\documentclass[]{aa}

\usepackage{xcolor}
\usepackage{graphicx}
\usepackage{txfonts}
\usepackage{float}
\usepackage{amsmath}
\usepackage{hyperref}
\usepackage{bm}
\usepackage{cases}

\newcommand\mueq{\mu_\mathrm{eq}}
\usepackage{hyperref}	% Hyperlinks
\hypersetup{colorlinks=true,linkcolor=blue,citecolor=blue,filecolor=blue,urlcolor=blue}
\begin{document}
\title{Spontaneous ring formation in wind-emitting accretion discs}
\titlerunning{Zonal flow instability in accretion discs}

\author{A. Riols\inst{1}\and  G. Lesur\inst{1}}
\institute{%
$^1$ Univ. Grenoble Alpes, CNRS, Institut de Planétologie et d’Astrophysique de Grenoble (IPAG), F-38000, Grenoble, France
}

\date{\today}

\abstract{Rings and gaps have been observed in a wide range of protoplanetary discs, from young systems like HLTau to older discs like  TW Hydra. Recent disc simulations have shown that  magnetohydrodynamic (MHD) turbulence (in the ideal or non-ideal regime) can lead to the formation of rings and be an alternative to the embedded planets scenario. In this paper, we investigate how these ring form in this context and seek a generic formation process,  taking into account the various dissipative regimes and magnetizations probed by the past simulations. We identify the existence of a linear and secular instability, driven by MHD winds, and giving birth to rings of gas having a width larger than the disc scale height. 
%An important prerequisite is the existence of a magnetically-driven wind whose intensity depends on the local disc magnetisation. 
%We perform 2D (axisymmetric) and 3D  MHD simulations to check the existence of the instability and bring evidence that the rings in these simulations are born through this mechanism. 
We show that the linear theory is able to make reliable predictions regarding the growth rates, ring/gap contrast and spacing, by comparing these predictions to a series of 2D (axisymmetric) and 3D MHD numerical simulations. In addition, we demonstrate that these rings can act as dust traps provided that the disc is sufficiently magnetised, with plasma beta lower than $10^4$. Given its robustness, the  process identified in this paper could have important implications, not only for protoplanetary discs but also for a wide range of accreting systems threaded by large-scale magnetic fields.}


\maketitle
%%%%%%%%%%%%%%%%%%%%%%%%%%%%%%%%%%%%%%%%%%%%%%%%%%

%%%%%%%%%%%%%%%%% BODY OF PAPER %%%%%%%%%%%%%%%%%%

\section{Introduction}
The radio-interferometer ALMA and the new generation of instruments like SPHERE at the Very Large Telescope have imaged a variety of structures in protoplanetary discs around young stars \citep{garufi17}. One of the most striking features are the concentric rings (or gaps), observed in many discs:  HL tau \citep{alma15},  TW Hydra \citep{andrews16} or the disc around Herbig Ae star HD 163296 \citep{isella16}.
These structures may influence  the disc evolution and could be a privileged location of dust accumulation \citep{pinilla12}, a key step towards planetary cores formation

One important challenge in accretion discs theory is to understand the origin of these rings. The scenario commonly invoked is the presence of embedded planets forming and opening gaps \citep{kley12,baruteau14,dong15}. Although the planet hypothesis is attracting and seems consistent with recent simulations  \citep{dipierro15},  it challenges the planet formation theory, in particular in young systems like HLTau (<1 Myrs old). The core accretion model at distance of a few tens of AU indeed requires more than a million years to form planets \citep{helled14}. A large number of alternative mechanisms have been suggested such as dust-drift-driven viscous ring instability \citep{wunsch05,dullemond18}, snow lines \citep{okuzumi16}, dead zones \citep{flock15}  or  secular gravitational instabilities in the dust \citep{takahashi14}. One recent and appealing scenario is the formation of concentric rings and gaps by magneto-hydrodynamics (MHD) processes in the disc. \\

Since the early 90s, it is admitted that MHD processes are ubiquitous  and crucial in the evolution of accreting systems.  Magnetized discs, if sufficiently ionized, are  indeed  prone to the magneto-rotational instability  \citep[MRI, ][]{balbus91,hawley95}, leading to turbulence and angular momentum transport.  When threaded by a mean vertical field, the disc may also evacuate a significant part of their angular momentum and energy through large-scale winds intimately connected to the MRI \citep{lesur13,fromang13}.  Even in poorly ionized regions ($r \gtrsim 0.1 - 1 AU$) subject to non ideal effects (ambipolar diffusion and Hall effect), accretion can operate via MHD winds, despite the absence of vigorous MRI turbulence \citep{bai13,lesur14,bai15,bethune17}. 

Several simulations in the local and global configuration, including a mean vertical field, have brought evidence that MHD flows and their winds, self-organise into large-scale axisymmetric structures  or "zonal flows" associated with rings of matter \citep{kunz13,bai15,bethune16,bethune17}. These features appear predominantly in the presence of non-ideal effects  but were also noticed in MRI simulations without any explicit diffusion \citep{steinacker02,bai14b,suriano18}.  Recent works attempted to explain their origin, though without any persuasive outcome.  \citet{bai14b} proposed that rings form through an "anti-diffusion" associated with the anisotropy of MRI turbulence. However, their result appears in conflict with most of the simulations that measured turbulent magnetic diffusivities  \citep{guan09,fromang09,lesur09b}. Using global simulations, \citet{suriano18, suriano18b} suggested that the structures are formed via reconnection of  pinched poloidal field lines in the midplane current layer. Nevertheless, there is a lack of evidence that this mechanism is generic and works in all magnetic configurations. It requires a particular symmetry, with a poloidal field bending in the midplane, which is not the geometry observed in many non-ideal simulations. \\

In this paper, we bring evidence that the process forming rings and gaps in MHD simulations (ideal, resistive or ambipolar) is generic and supported by a local wind instability. The instability requires a mean mass ejection and a radial transport of vertical magnetic flux, whose origin can be the "$\alpha$" viscosity or the zonal flow itself.  In turbulent discs, the criterion for instability is that the mass loss rate increases faster than the stress with the disc vertical magnetization. In presence of turbulence, the mechanism is reminiscent of a "viscous-type" instability, like imagined by \citet{lightman74}, except that the mass is free to escape the disc vertically. In a sense, it also shares some similitude with the wind-driven instability proposed by \citet{lubow94} and \cite{cao02}.  Unlike the latter, however,  it does not rely on the assumptions that angular momentum is removed by the wind magnetic torque, neither that the mass loss rate increases with the poloidal field inclination, which are arguable assumptions \citep[see ][]{konigl96}. We believe that the mechanism described in this paper is closely related to the "mass-flux" or "stripe" instability seen by \citet{moll12} and \citet{lesur13}  in the highly magnetized (MRI-stable) regime. 

The plan of the paper is as follows:  in Section \ref{sec_phenomenlogy}, we present the main characteristics of zonal flows (or rings) in MHD simulations and  show that unlike the intuitive sense, they do not form through a radial transport of matter,  but appear as a consequence of gaps emptied by vertical outflows. In Section \ref{sec_theory}, we suggest the existence of a wind-driven instability and calculate its growth rate theoretically. In Section \ref{sec_simulations}, we perform 2D MHD simulations (with or without explicit diffusion)  for a wide range of magnetizations to test the existence and properties of the instability. We show in particular that axisymmetric modes projected into the Fourier space grow exponentially, with well-defined growth rates corresponding to those predicted by the theory. We also explore the nonlinear saturation of the instability and attempt to predict the rings/gap contrast and their radial separation. Finally, in section \ref{sec_discussions}, we discuss the potential implications of our work beyond the scope of stellar discs. 

\section{Phenomenology of self-organization}
\label{sec_phenomenlogy}
\subsection{Zonal flows and rings occurrence in MHD simulations}
\begin{figure}
\centering
\includegraphics[width=\columnwidth,trim={2.2cm 8cm 2cm 0cm},clip]{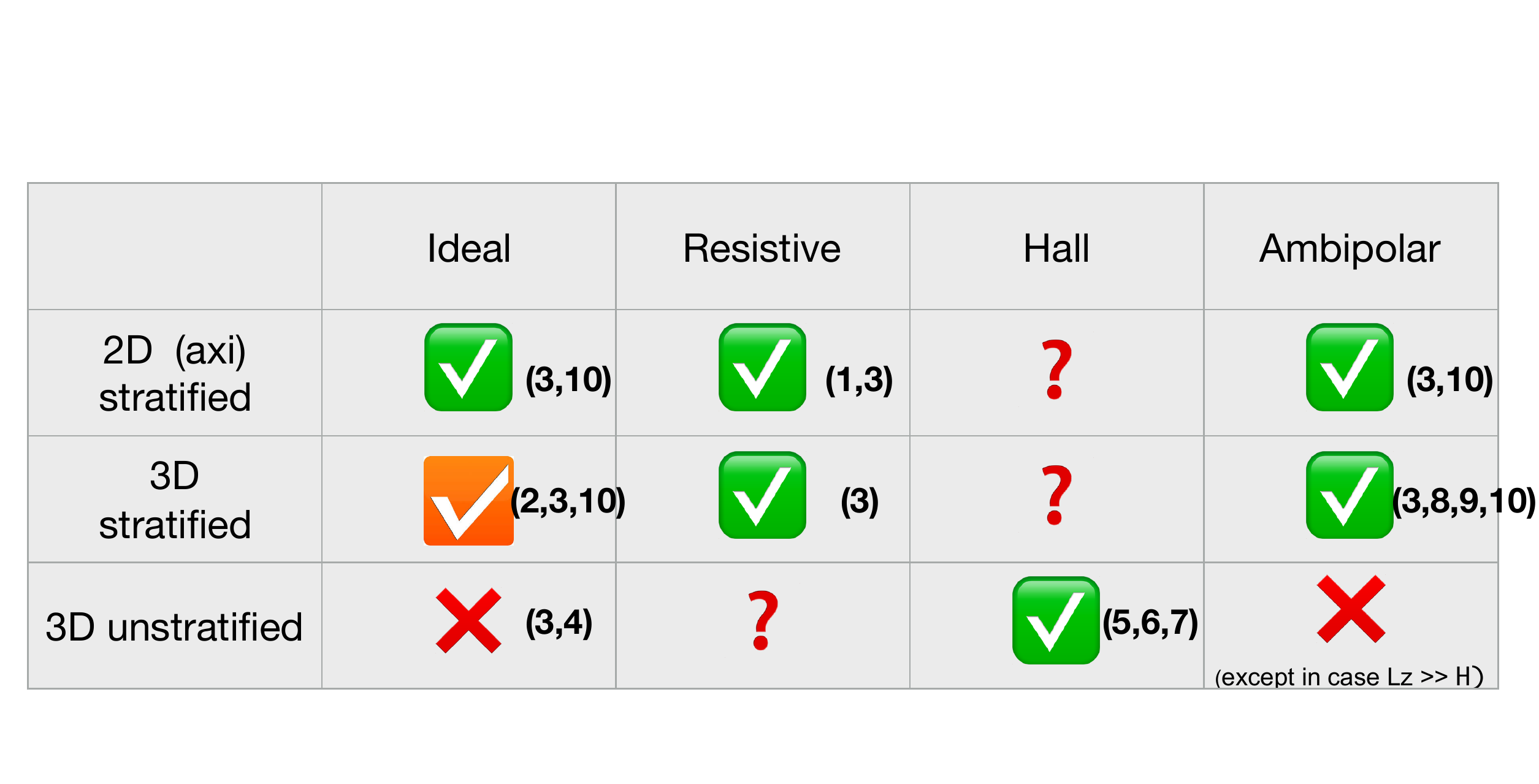}
 \caption{Ring occurence in MHD simulations with net vertical magnetic flux. The orange check mark means that zonal flow exist, but their persistence within the turbulent flow and their convergence with box size are uncertain at current time. (1) \citet{suriano17}, (2)  \citet{bai14b}, (3) our own simulations, (4) \citet{hawley01}, (5) \citet{kunz13}, (6) \citet{bethune16}, (7) \citet{krapp18}, (8) \citet{bai15}, (9) \citet{riols18}, (10) \citet{suriano18,suriano18b}}
\label{fig_occurence}
\end{figure}
Self-organization of the gas  into "zonal flows" is found in various MHD simulations of discs threaded by a net vertical field. The term "zonal flows" refers to the succession of sub-keplerian and super-keplerian bands,  associated with large scale rings of matter, as revealed by the global simulations of \citet{bethune17}. In the following, the term "ring" or "zonal flow" will be used to refer to the same sub-structure. Fig.~\ref{fig_occurence} summarizes ten years of research and numerical effort to identify the conditions under which zonal flows can develop.

In 3D unstratified simulations, zonal structures seem absent in the ideal or ambipolar cases but arise in the Hall-dominated regime via an anti-diffusive process \citep{kunz13,bethune16,krapp18}.  Early global simulations without explicit diffusion and neglecting the vertical gravity \citep{hawley01,steinacker02} have yet reported gaps and rings,  whose origin were originally attributed to a "viscous-type" instability. However, the instability criterion is generally not fulfilled in  simulations (see Section \ref{stability_criterion}), and these structures may be created through a boundary effect (perhaps artificial) relying on the pile-up of matter at the  disc inner edge.  Our own local unstratified simulations in the ideal limit or with ambipolar diffusion (not shown here) indicate that self-organized structures vanish as the horizontal box size is extended. Such behaviour is not obtained in stratified simulations, for which the rings separation converges with the radial box size (see Section \ref{sec_simulations3D}). \\

Stratified simulations (2D or 3D) show a radically different result:,  ambipolar diffusion \citep[with or without ohmic diffusion, see][]{simon14,bai15,riols18,suriano18} and plasma combining the three non-ideal effects \citep[Ohmic, Hall and ambipolar, see][]{bai15} favour the emergence of zonal flows. However, in stratified discs, it is still uncertain whether the Hall effect alone can trigger their formation  \citep{kunz13,lesur14}. Although ambipolar diffusion is believed to enhance the process of rings formation,  local and global simulations without explicit diffusion \citep{bai14b,suriano18b} or with pure ohmic resistivity  \citep[][see also Section \ref{sec_simulations}]{suriano17}  also exhibit an efficient production of large-scale rings. This is particularly obvious in the 2D axisymmetric case when fields are independent of the azimuthal coordinate.  Perhaps, the 3D ideal case is a matter of discussion, since the structures obtained are generally more difficult to identify and fill the box entirely in local simulations. Actually, we will see in Section \ref{sec_simulations3D} that rings of intermediate size undeniably form in 3D,  but are efficiently diffused by vigorous non-axisymmetric MRI turbulence. \\

Note that axisymmetric structures have been also discovered in zero net flux simulations \citep{johansen09,simon12} but are transient and emerge probably from stochastic processes.  If we put aside the Hall-effect, whose role in rings formation is probably of different nature \citep{kunz13}, and seek for a generic mechanism,  it comes naturally from Fig.~\ref{fig_occurence} that the rings formation process is ideal in essence and requires a large scale poloidal field with a vertical stratification. 
\begin{figure}
\centering
\includegraphics[width=\columnwidth]{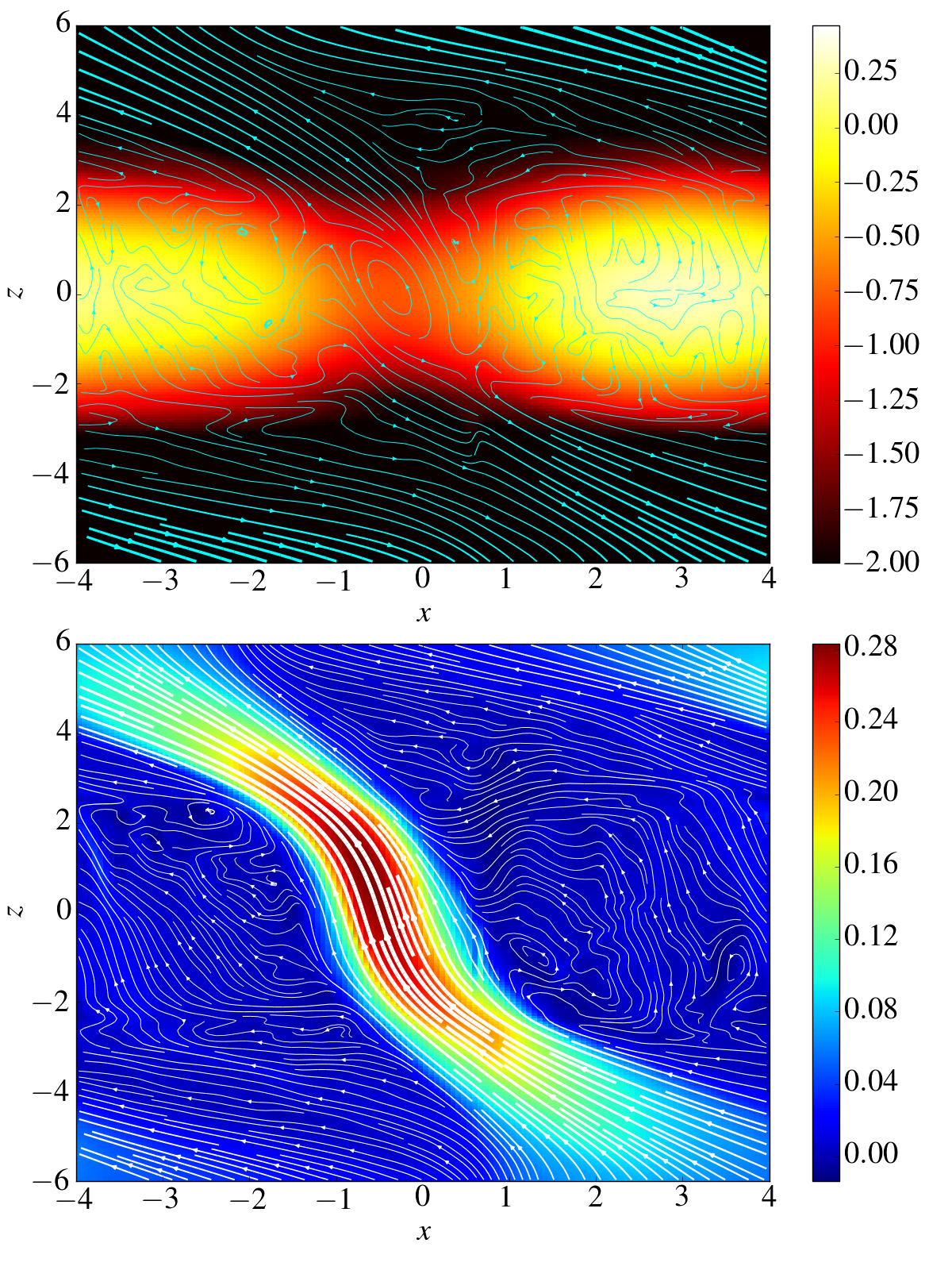}
 \caption{Example of a ring/gap structure and its surrounding outflow topology in the 3D shearing box simulation of \citet{riols18} with ambipolar diffusion. Top panel: gas density (colormap in log scale) and  streamlines (cyan lines) in the poloidal plane  ($x$ and $z$ are respectively  the radial and vertical coordinates).  Bottom panel:  vertical magnetic field (colormap) and poloidal magnetic field lines (in white)  showing the  inclined "plume" structure from where most of the mass is extracted. The thickness of the lines is proportional to the field  intensity. }
\label{fig_ring_beta3Am1}
\end{figure}
\begin{figure}
\centering
\includegraphics[width=\columnwidth]{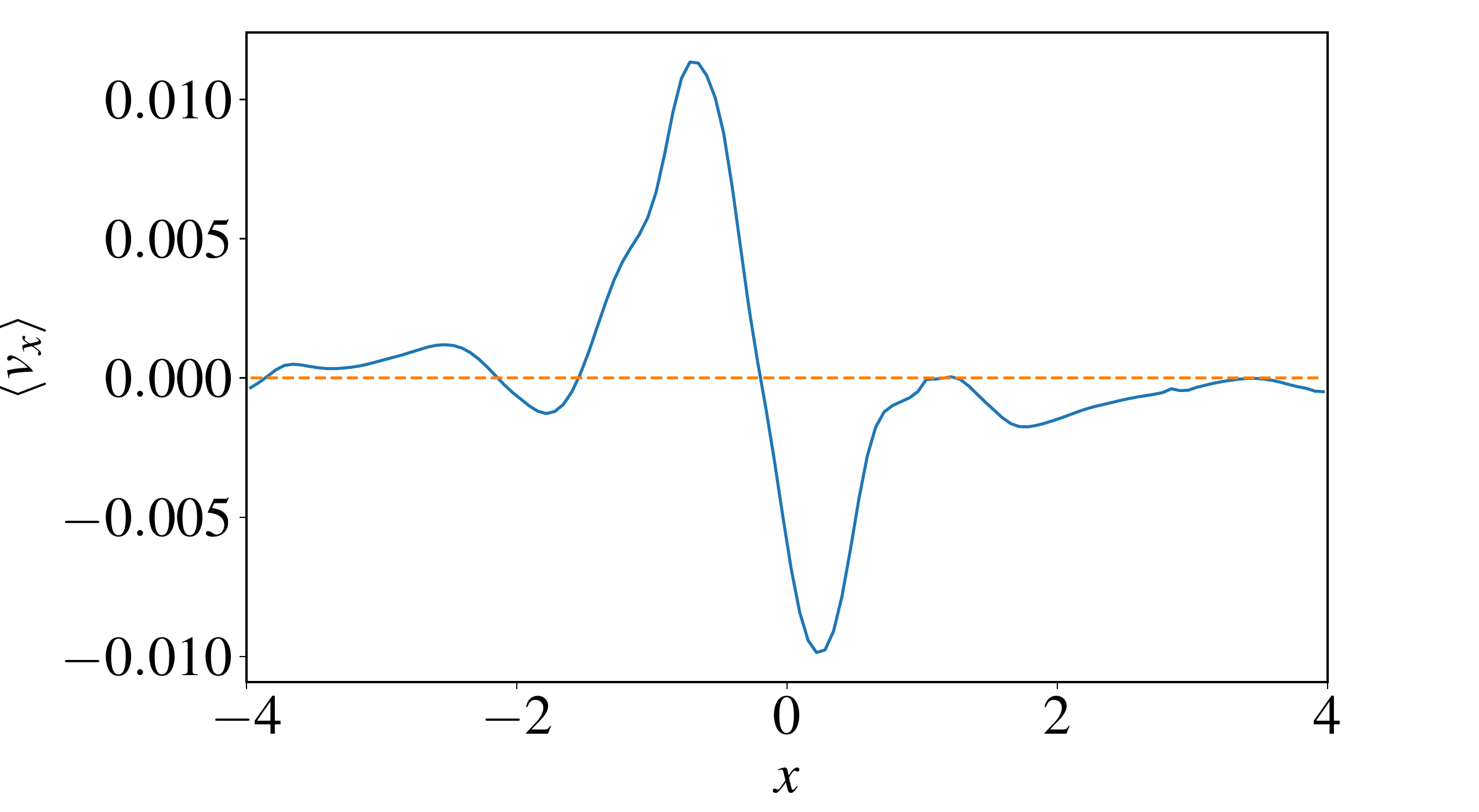}
\caption{Radial velocity computed from the 3D simulation of Fig.~\ref{fig_ring_beta3Am1}. It is averaged in the azimuthal direction, in $z$ within  $\pm 1.5 H$ and during the first 50 orbits ($\sim300 \,\Omega^{-1}$), corresponding to the growth of the ring structure.}
\label{fig_vxmean}
\end{figure}
\begin{figure}
\centering
\includegraphics[width=\columnwidth,trim={2.2cm 0 2cm 0},clip]{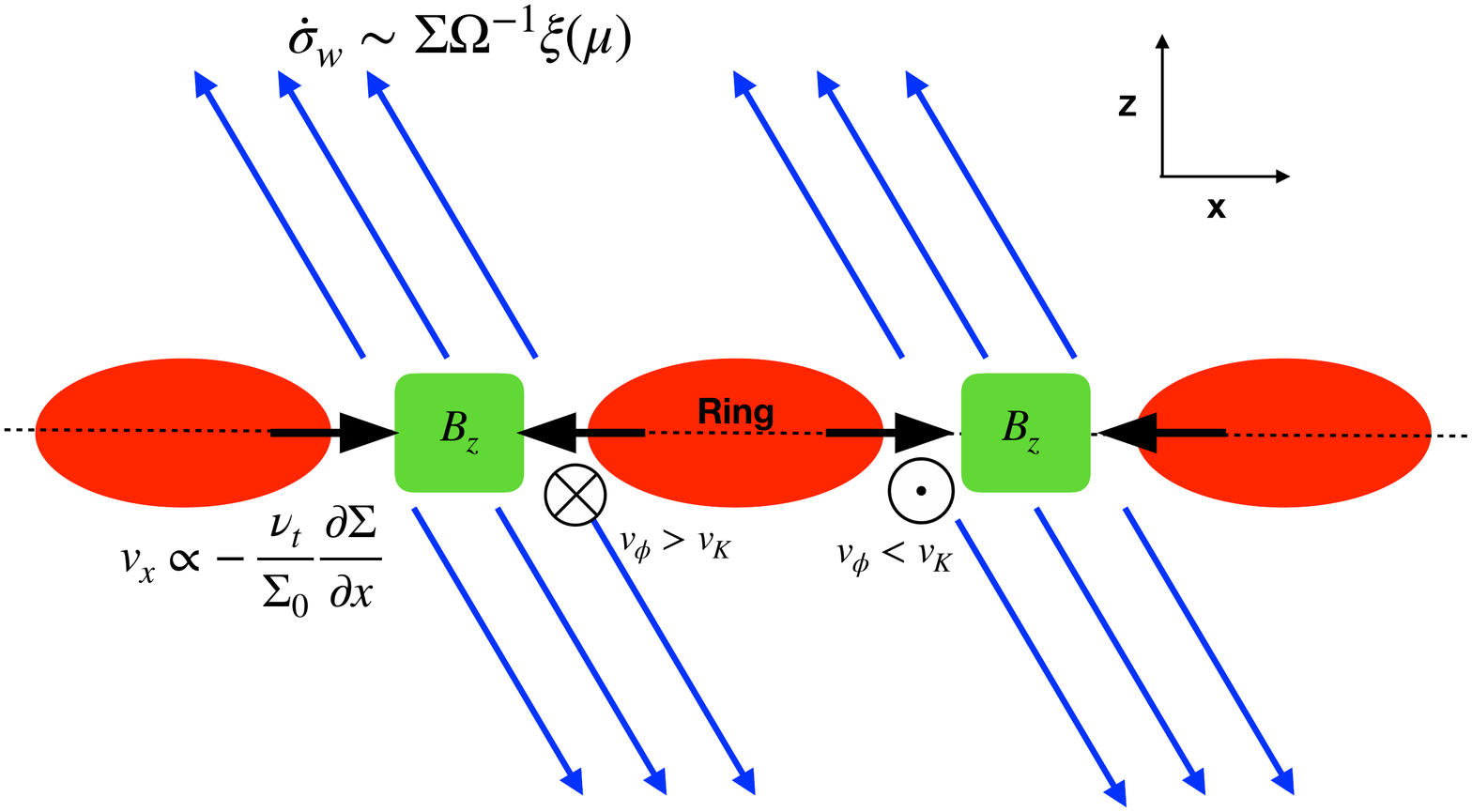}
 \caption{Sketch illustrating the development of rings in accretion discs. The black and blue arrows represent respectively the radial flow and  the wind.  The green patches correspond to density minima where the vertical field $B_z$ and the magnetization $\mu$ grow. In this representation, the poloidal field lines (// to the wind) do not bend in the midplane but cross the disc with some angle. This  configuration is actually not intuitive but is encountered in various disc simulations around the midplane.}
\label{fig_sketch_instability}
\end{figure}
\subsection{Characteristic features}
We remind here the generic characteristics of zonal flows found in various MHD simulations (Hall free regime). Figure \ref{fig_ring_beta3Am1}  shows the typical flow and magnetic field topology around such structure, obtained in the 3D simulation of \citet{riols18} with ambipolar diffusion.  The quantities are averaged in time and in the azimuthal direction, for a magnetization $\mu=10^{-3}$ and ambipolar coefficient $Am=1$ in the midplane  \citep[see][for more details about the numerical setup]{riols18}. We identify in the top panel a strong and coherent windy plume that emanates from the density minimum (gap). The structure is inclined and connected to a large scale roll in the midplane.  The bottom panel shows that the poloidal magnetic field is concentrated radially within the gap, while outside the net vertical flux is close to zero. It follows that the  magnetization (defined as the inverse of the plasma beta-parameter in the midplane)
\begin{equation*}
\mu = \beta^{-1} =\dfrac{B_z^2}{2\rho_0 c_s^2},
\end{equation*}
with $\rho_0 ={\Sigma}/{\sqrt{2\pi}}$,  is  stronger in the gaps and therefore anti-correlated with the rings \citep[see also][for instance]{suriano18}. The poloidal field lines follow the plume and clearly drive the outflow in this region.  The existence of such localized and inclined magnetic shell is reminiscent of other MHD simulations \citep[ideal or not, see][]{bai14b,bai15} and their origin seems clearly connected to the formation of the ring.  Note that the inclined outflow and the spontaneous breaking of vertical symmetry is  found in various simulations, including global models \citep[Fig.~25 of][]{bethune17,gressel15b}. Thus they cannot be considered as glitches of the shearing box. We stress however that local models do not correctly reproduce the outflow behaviour at large distance from the disc midplane. Actually, if the star is located on the left in Fig.~\ref{fig_ring_beta3Am1}, the upper field lines must bend at some point, further out in the atmosphere (see Fig.~25 of \citet{bethune17}).
\subsection{The key role of wind plumes}
Another important and recurrent feature of the ring structure is related to the radial velocity profile.  Fig.~\ref{fig_vxmean} shows that the averaged $v_x$  within the midplane ($\vert z \vert \lesssim H$)  is opposed to the formation of the ring. In particular, it is anti-correlated with the azimuthal velocity perturbation (zonal flow) and phased with an angle of $\pi/2$ with respect to the density maximum. During the ring formation, the matter is thus radially dragged toward the gap in average. Physically, this is expected since the turbulent stress acts like an effective viscosity that tends to diffuse any density bulge in the disc. Therefore, the only way to produce a ring is through vertical transport of matter. We found that the strong wind plume identified in Fig.~\ref{fig_ring_beta3Am1} indeed clear out the gas in the gap regions. In other words, the ring structure does not result from matter being radially concentrated  but from material in the gaps being depleted. Note that a similar result is obtained for other magnetizations and in larger box simulations with more than one ring. A detailed mass budget is provided in Appendix \ref{appendixD} for such simulation. In the next section, we show that a wind-driven instability is probably connected with the formation of these plumes, which organize the flow into radial sub-structures.
\section{A unified theory based on a wind instability}

\label{sec_theory}
\subsection{Naive picture of the instability}
We set here the basic principles of a wind instability, that may lead to the growth of ring structures in accretion discs. To start with, consider a classical "$\alpha$" disc with constant and uniform viscosity $\nu_t = \alpha c_s H$.  Assume initially a small axisymmetric and radial perturbation of the surface density $\delta{\Sigma}$. In the local approximation, then the viscous stress generates a radial flow associated with angular momentum transport,
\begin{equation}
\label{eq_viscous_accretion}
\delta{v}_R = -\dfrac{2\nu_t}{\Sigma} \dfrac{\partial \,\delta {\Sigma}}{\partial R},
\end{equation}
which acts to diffuse the initial density ring perturbation. This is a classical and trivial result of the viscous disc theory.  In the absence of winds,  the perturbation decays and the disc remains stable.  If now we include a wind, associated with a large-scale poloidal magnetic field that removes the mass and angular momentum, the outcome can be radically different and potentially lead to an instability.  Figure \ref{fig_sketch_instability} sketches out the main dynamical
ingredients of such instability. If the magnetic diffusivity is not too high, the radial flow generated by the viscous stress concentrates the vertical magnetic flux $B_z$ in the density minima or gaps (green patches). Because matter and magnetic flux are transported radially at the same speed $\delta v_R$,  we expect the magnetization $\mu \propto  {B_z^2}/({\Sigma c_s \Omega})$ to increase within the density minima. In such circumstances, the wind originally uniformly distributed in $R$, will adjust to this new configuration. A key hypothesis is  that the vertical mass flux of the wind $\dot{\sigma}_w$ (proportional to the surface density)  increases with $\mu$.   Therefore,  if the magnetized gaps eject more material than viscosity can bring in, the initial perturbation is reinforced. 

Quantitatively, an instability occurs whenever the surplus of mass launched in the wind is larger than the mass flowing radially. If we note $r_a$ the typical separation between the rings and $p=d \log \dot{\sigma}_w/ d \mu>0$,   the instability criterion becomes: 
\begin{equation*}
r_a \dot{\sigma}_w \left[2p  \dfrac{\delta {B}_z}{B_z} - (p-1) \dfrac{\delta{\Sigma}}{\Sigma}\right] >  \Sigma \,\delta v_R.
\end{equation*}
Using relation (\ref{eq_viscous_accretion}) and the transport equation for $B_z$, it is possible to show that the instability criterion is simply $p>0$.  In sum, the instability is driven by the combination of an outflow and a radial flux transport, and appears optimal in the ideal limit. Rings do not originate from a radial concentration of matter but result from the vertical depletion of their surrounding gaps. The instability is of same nature as \citet{lubow94} and \citet{cao02} but does not require a wind torque (this point is discussed in Section \ref{sec_discussions}). Note that the radial pressure gradient associated with the rings has to be balanced by the Coriolis force. The geostrophic equilibrium naturally gives birth to a zonal flow with $v_\phi > v_K$ in the regions where $v_R<0$ and $v_\phi < v_K$ where $v_R>0$ (see Figure \ref{fig_sketch_instability}). 

Although the instability mechanism relies on simple physical arguments,  it needs to be demonstrated rigorously through a linear analysis of the MHD equations. Moreover there are many caveats to the simple picture described here: first the viscosity  or transport coefficient $\alpha$  generally depends on $\mu$,  which is a widely accepted result based on turbulent MHD simulations.  This effect may reduce the strength of the instability.   Second, what happens if angular momentum is free to flow along the poloidal magnetic field line? The presence of a mean toroidal field and a vertical stress  could in principle have important consequences on the re-distribution of angular momentum. Finally, how does the instability behave if the disc is subject to magnetic diffusion and non-ideal effects? In the next section, we derive a general instability criterion taking into account several of these effects. 
\subsection{Averaged equations in the local framework}
\label{averaged_equations}
To simplify the problem, we use the local shearing sheet framework \citep{goldreich65}. This corresponds to a Cartesian patch of the disc, centred at $r=R_0$, where the Keplerian rotation is approximated locally by a linear shear flow plus a uniform rotation rate $\boldsymbol{\Omega}=\Omega \, \mathbf{e}_z$. We note  $(x,y,z)$ respectively the radial, azimuthal and vertical directions. 

To analyse the  radial disc structure, we integrate azimuthally and vertically the equations of motion and therefore neglect the vertical dependence of the flow.  For that purpose, we introduce two average procedures in the plane $(y,z)$: 
a standard average $\langle \underline{\cdot} \rangle$  between $-z_d$ and $z_d$, where $z_d$ is some arbitrary altitude, and a mass-weighted average $\overline{\,\cdot\,}$ so that for any field $\varphi$, we have: 
\begin{equation*}
\langle \varphi \rangle(x) = \dfrac{1}{L_y}\int\int_{-z_d}^{z_d} \varphi \, dz \,dy  \quad \text{and} \quad  \langle \underline{\varphi}\rangle =  \langle \varphi \rangle / (2 z_d),
\end{equation*}
\begin{equation*}
\overline{\varphi}(x)  = \dfrac{1}{L_y \Sigma}\int\int_{-z_d}^{z_d} \rho  \varphi \, dz\,dy,
\end{equation*}
where $\rho$ is the fluid density and $\Sigma(x) = \langle \rho \rangle = \int  \int_{-z_d}^{z_d} \rho\,  dz dy$  the surface density. Each field can be decomposed  into a sum of a mean component (depending on $x$ only) and a fluctuation 
\begin{equation}
\label{eq_decomposition}
{\varphi} = \overline{\varphi}(x) + \varphi'(x,y,z)  \quad \text{or} \quad {\varphi} = \langle \underline{\varphi}\rangle(x)  + \delta \varphi(x,y,z).
\end{equation}
We also note $[\dot]^{+}_-$ the difference between the field at $z=z_d$ and $z=-z_d$.  The compressible, inviscid and  isothermal equations of motion in the horizontal plane, integrated azimuthally and vertically between $-z_d$ and $z_d$ are: 
\begin{equation}
\dfrac{\partial {\Sigma}}{\partial t}+\dfrac{\partial}{\partial x}\left(\Sigma \overline{v}_x\right)+ \dot{\sigma}_w=\dot{\sigma}_i,
\label{mass_eq}
\end{equation}
\begin{equation}
\dfrac{\partial {\Sigma \overline{v}_x}}{\partial t}+\dfrac{\partial}{\partial x}\left(\Sigma\overline{T}_{xx}\right)+{W}_{xz}-2\Omega \Sigma \overline{v}_y+c_s^2\dfrac{\partial \Sigma}{\partial x}+\dfrac{\partial} {\partial x} \dfrac{\langle{B^2\rangle}}{2}=0,
\label{mx_eq}
\end{equation}
\begin{equation}
\dfrac{\partial {\Sigma \overline{v}_y}}{\partial t}+\dfrac{\partial}{\partial x}\left(\Sigma \overline{T}_{yx} \right)+{W}_{yz}+\dfrac{1}{2}\Omega \Sigma \overline{v}_x=0,
\label{my_eq}
\end{equation}
where 
$\dot{\sigma}_w  = \left [\rho v_z\right]_{-}^{+}$ is the mass loss rate,  $\overline{T}_{ij}  =\overline{v_iv_j}-\overline{{B_iB_j}/{\rho}}$ the stress tensor integrated in the vertical direction and $W_{ij} = \left[\rho v_i v_j - B_iB_j\right]_{-}^{+}$  its boundary value. We assume that mass is replenished locally at a constant rate $\dot{\sigma}_i$ (associated with the radial accretion in a global view, which is absent in the local shearing box framework). \\

These equations of motions are coupled with the induction equation that describes the evolution of the magnetic field $B$. We are particularly interested in the evolution of the vertical magnetic flux $\langle B_z\rangle$:  
\begin{equation}
\dfrac{\partial \langle {B}_z \rangle }{\partial t}=\dfrac{\partial}{\partial x}\langle  \mathcal{E}_y \rangle+ \eta  \left (\dfrac{\partial^2}{\partial x^2}\langle B_z\rangle+ \left[\dfrac{\partial B_z}{\partial z}\right]_{-}^{+}\right),
\label{bz_eq}
\end{equation}
where $\mathcal{E}_y ={v}_z  B_x - {v}_x B_z$ is the toroidal electromotive force and $\eta$ is assumed to be a constant and uniform ohmic resistivity. An additional constraint is the solenoidal condition which gives:
\begin{equation*}
\dfrac{\partial \langle B_x \rangle}{\partial x}+\left[ B_z\right]^+_-=0
\end{equation*}
\subsection{Stress tensor and electromotive force}
We first develop the terms related to the stress tensor  $\overline{T}_{ij}$ and $W_{ij}$.  In the limit of highly subsonic fluctuations,  it is straightforward to show that the radial stress $\Sigma \overline{T}_{xx}$ and the vertical stress in the radial momentum equations are negligible compared to thermal pressure. Thus we can assume that $\overline{T}_{xx}\simeq 0$ and $W_{xz}\simeq 0$. In the azimuthal momentum equation, however, the stress is comparable to others terms. By using the decomposition of Eq.~(\ref{eq_decomposition}), we have 
\begin{equation}
\label{eq_Txy}
\Sigma \overline{T}_{yx} =\Sigma \overline{v_xv_y}-{\langle{B_xB_y}\rangle}= (\alpha_\nu + \alpha_L) \Sigma c_s^2 
\end{equation}
where 
\begin{equation*}
\alpha_\nu = \dfrac{\langle \rho v_x'v_y'-\delta B_x\, \delta B_y\rangle}{\Sigma c_s^2}  \quad \text{and}    \quad \alpha_L = \dfrac{\Sigma \overline{v_x} \,\overline{v_y}-{\langle{B_x} \rangle \langle \underline{B_y}\rangle}}{\Sigma c_s^2}
\end{equation*}
can be identified respectively as a turbulent and laminar radial transport coefficient. The term related to the vertical stress $W_{yz}$ can be written as: 
\begin{equation}
W_{yz} =  \alpha_W  \Sigma c_s \Omega  + \overline{v}_z\left[v_y'\right]_{-}^{+}+ \overline{v}_y\left[v_z'\right]_{-}^{+}  - \langle \underline{B_{y}}\rangle\left[ \delta B_z\right]_{-}^{+} - \langle \underline{B_{z}} \rangle\left[ \delta B_y\right]_{-}^{+} 
\label{eq_Wyz}
\end{equation}
where  $\alpha_W=\dfrac{\left[\rho v_y' v_z' - \delta B_y\,\delta B_z\right]_{-}^{+}}{\Sigma c_s \Omega}$ is the turbulent vertical transport. Finally the electromotive force in Eq.~(\ref{bz_eq}) can be decomposed as well into a laminar and a turbulent part, the latter being assumed to behave as an effective magnetic diffusivity $\eta_t$ : 
\begin{equation*}
\langle \mathcal{E}_y\rangle = \langle\underline{ v_z} \rangle \langle B_x \rangle -  \langle  \underline{ v_x} \rangle \langle B_z \rangle + \eta_t  \dfrac{\partial \langle B_z \rangle }{\partial x}
\end{equation*}

\subsection{Power laws for mass loss rate and turbulent coefficients}
\label{closure_relations}
To close the system of equations we need to relate  the mass loss efficiency $\zeta = \dot{\sigma_w}/(\Sigma \Omega^{-1})$ and the turbulent coefficient $\alpha_\nu$, $\alpha_W$,  $\eta_t$ to the integrated disc quantities. It is reasonable to assume that these coefficients depend principally on the main dimensionless parameter of the disc, namely the vertical magnetization $\mu$. We suppose that this dependence can be captured by a simple power law:
\begin{equation*}
\zeta  =\dfrac{\dot{\sigma_w}}{\Sigma \Omega^{-1}} =   \zeta_0\left(\dfrac{\mu}{\mueq} \right)^p
\end{equation*}
\begin{equation*}
\alpha_\nu= \alpha_{{\nu}_0}  \left(\dfrac{\mu}{\mueq} \right)^{q} 
\end{equation*}
and similar relations for $\alpha_W$ and $\eta_t$.  $\mueq$, $\zeta_0$ and $\alpha_{{\nu}_0}$ correspond to values of a given equilibrium (see next section). Numerical simulations are actually a suitable tool to probe and test these scaling laws. The dependence of $\alpha_\nu$ on the magnetization has been explored in various ideal simulations of MRI turbulence, with or without magnetic diffusion and thermal effects  \citep{hawley95,simon13,salvesen16,scepi18}. In all cases, there is  some consensus that 
\begin{equation*}
q \simeq 0.5  \quad \text{for} \quad  \mu \gtrsim 10^{-5} 
\end{equation*}
The vertical transport due to a wind has also been measured in simulations \citep{bai13,fromang13,zhu18,scepi18} but is generally supplied by a large scale toroidal field instead of turbulent fields. Characterizing properly $\alpha_W$ would require to measure the laminar and turbulent contribution of the stress  separately, which has never been done in the literature. The turbulent diffusivity has been calculated in 3D unstratified shearing box simulations \citep{guan09,fromang09,lesur09b} and a fair assumption is to consider $\eta_t$ between 0.2 and $0.5\nu_t$ (where  $\nu_t =\alpha_\nu c_s \Omega$). We will see in Section \ref{sec_simulations} however that such ratio can be actually much weaker in 2D. Finally, evaluating the mass loss rate is probably the hardest part since it  depends on the vertical box size and the nature of boundary conditions \citep{fromang13}.  Simulations of 1D laminar winds  predict $p\approx 0.6- 0.7$ \citep[see Fig.~4 of][]{riols16b} while full 3D turbulent simulations suggest $p \simeq 1$; see Fig.~5 of \citet{suzuki09} and Fig.~4 of \citet{scepi18}. 
   %W_{yz}= \alpha_{w} \Sigma c_s^2/H + i   = \dfrac{ \alpha_{{w}_0}}{H}  \left(\dfrac{\mu}{\mueq} \right)^{W_\mu} \Sigma c_s^2

\subsection{Local equilibrium solutions}
We note with a subscript "0" the equilibrium solutions  (independent of time and $x$) of the system of equations derived in  Section \ref{averaged_equations}. The local disc equilibria are obtained by setting ${\partial {\cdot }}/{\partial t}=0$ and ${\partial {\cdot }}/{\partial x}=0$ in Eqs.~\ref{mass_eq} \ref{mx_eq}, \ref{my_eq} and \ref{bz_eq}.  Note that in absence of turbulence, these equilibria correspond to the vertical average of the 1D (z-dependent) wind solutions studied by \citet{lesur13} and \citet{riols16b}. The solenoidal condition gives immediately  $\delta B_z=0$, which means that $B_z$ = $\langle \underline{B_z} \rangle = B_{z_0}$. We can then define a constant magnetization of the equilibrium,  as: 
\begin{equation*}
\mueq = \sqrt{\dfrac{\pi}{2}} \dfrac{ B_{z_0}^2 }{\Sigma_0 c_s \Omega}  
\end{equation*}

 Solutions can be either symmetric about the midplane with $B_x= B_y =0$ at $z=0$  or antisymmetric with $\partial_z B_x=\partial_z B_y=0$. 
\begin{enumerate}
\item In the symmetric case, we have $\langle B_x \rangle = \langle B_y \rangle =0 $ but $\overline{v}_x,\overline{v}_y   \neq 0$.  Horizontal components satisfy $\left[v' \right]_{-}^{+}=0$ but $\left[\delta b \right]_{-}^{+}\neq 0$. Therefore there is a net vertical stress through the disc, which is responsible for a mean accreting flow.  \\

\item  In the second case  $\langle B_x \rangle = \langle B_y \rangle \neq 0 $ but $\overline{v}_x,\overline{v}_y  = 0$.  We define $B_{x_0}$ and $B_{y_0}$  the mean radial and toroidal magnetic field throughout the midplane. Departures to the vertical average satisfy  $[v_{x,y}']_{-}^{+} \neq 0$ but $\left[\delta b \right]_{-}^{+} = 0$.  It is straightforward to check that $W_{{yz}_0}=0$ when such symmetry is enforced. 
\end{enumerate}
 
Historically, the first class of solutions (1) were considered as the most intuitive and representative of a disc structure; the reason being that magnetic field lines do not bend outside the midplane.  However antisymmetric solutions have been shown to  naturally emerge from turbulent shearing box simulations \citep{lesur14,bai15}, and even from global simulations when non-ideal effects are included \citep{bethune17,bai17}. For that reason, we will mainly focus on the antisymmetric solutions in the next sections. 
\subsection{Linearisation around equilibrium}
\label{linearisation}
Let us note with the subscript $``0''$ an equilibrium solution of Eqs.~\ref{mass_eq} \ref{mx_eq}, \ref{my_eq} and \ref{bz_eq} enforcing the second class of symmetry. We remind that such symmetry implies that $\overline{v}_{0} = 0$ at equilibrium. To study the stability of these solutions, we introduce small normalized axisymmetric perturbations of the form $\hat{\varphi} \propto \exp{(ik_x x+\sigma t)}$ around the equilibrium flow:
\begin{equation*}
\Sigma = \Sigma_0 (1+ \hat{\Sigma}); \quad \overline{v_x} = \hat{u} c_s; \quad  \overline{v}_y = \hat{v} c_s ; \quad  \langle B \rangle= B_{0}  (1 + \hat{b})
\end{equation*}. We can do a similar decomposition for the magnetization 
\begin{equation*}
\mu = \mueq(1+\hat{\mu}) =  \mueq(1+2\hat{b}_z - \hat{\Sigma}) \\
\end{equation*}
as well as the turbulent coefficients and mass loss rate whose first order perturbation is proportional to $\hat{\mu}$. To be consistent with the formalism of Section \ref{averaged_equations}, the radial scale of perturbations is supposed to be much larger than the turbulent lengthscale. Because turbulent eddies are in principle limited to the disc scaleheight, we consider modes with $k_x H \lesssim 1$ and  assume $\sigma \ll \Omega$.  To simplify the problem, we make further assumptions: 

\begin{enumerate}
\item Perturbations are in a geostrophic equilibrium (pressure gradient balances Coriolis force in the $x$ direction). This is satisfied if $\sigma\ll \Omega$. \\
\item The vertical component of the Reynolds stress tensor   $\overline{v}_z\left[v_y'\right]_{-}^{+}+ \overline{v}_y\left[v_z'\right]_{-}^{+}$ as well as the turbulent vertical stress $\alpha_W=0$ are neglected \\
\item The  stress is purely turbulent  $\alpha_{\nu_0} \gg \alpha_{\L_0}$ \\
\item In Eq.~(\ref{bz_eq}), the vertical stretching of radial field $\langle u_z \rangle\langle B_x \rangle$ and the vertical diffusion of $B_z$ are neglected.  \\
\item We assume that $\langle \underline{v_x} \rangle \simeq \overline{v}_x$. This is true if the radial velocity perturbation does not vary too much between  -$z_d$  and $z_d$. \\
\end{enumerate}
Some of these assumptions are tested in simulations (see Appendix  \ref{appendixC} and \ref{appendixD}).  A more general case, including a laminar stress and a departure from the geostrophic equilibrium due to a strong toroidal field, is treated in  Appendix \ref{appendixB}.  By using these hypothesis, we show that the linearised momentum transport due to the total  stress at leading order simply reduced to
\begin{equation*}
\dfrac{\partial}{\partial x}\left ({\Sigma \overline{T}_{yx}}\right)+ {W}_{yz} = i k_x \alpha_{\nu_0}  \left [  \hat{\Sigma} +  q \hat{\mu}\right]  \Sigma_0  c_s^2
\end{equation*}
Note that the perturbation of  vertical stress  cancels out with the term -$\langle B_y \rangle \langle \underline{B_x}  \rangle \hat{b}_x$  in the azimuthal momentum equation (see Appendix \ref{appendixA}).  We normalize $k_x$  to $H$ and $\sigma$ to $\Omega$  and define $\eta^\star=(\eta + \eta_0)/(H^2 \Omega)$.  Linearisation of equations (\ref{mass_eq}), (\ref{mx_eq}), (\ref{my_eq}) and (\ref{bz_eq})  around equilibrium leads to:
\begin{equation}
\sigma \hat{\Sigma}+ ik_x  \hat{u} + p \zeta_0 \hat{\mu} + \zeta_0 \hat{\Sigma}=0
\label{mass_eq_lin}
\end{equation}
\begin{equation}
 - 2 \hat{v} +ik_x \hat{\Sigma}=0,
\label{mx_eq_lin}
\end{equation}
\begin{equation}
\sigma \hat{v}\,+ ik_x \alpha_{\nu_0} \left(q \hat{\mu} + \hat{\Sigma}\right) +\dfrac{1}{2} \hat{u}=0,
\label{my_eq_lin}
\end{equation}
\begin{equation}
\sigma \hat{b}_z+ ik_x \hat{u}= - \eta^\star k_x^2 \hat{b}_z,
\label{bz_eq_lin}
\end{equation}
\begin{figure}
\centering
\includegraphics[width=\columnwidth]{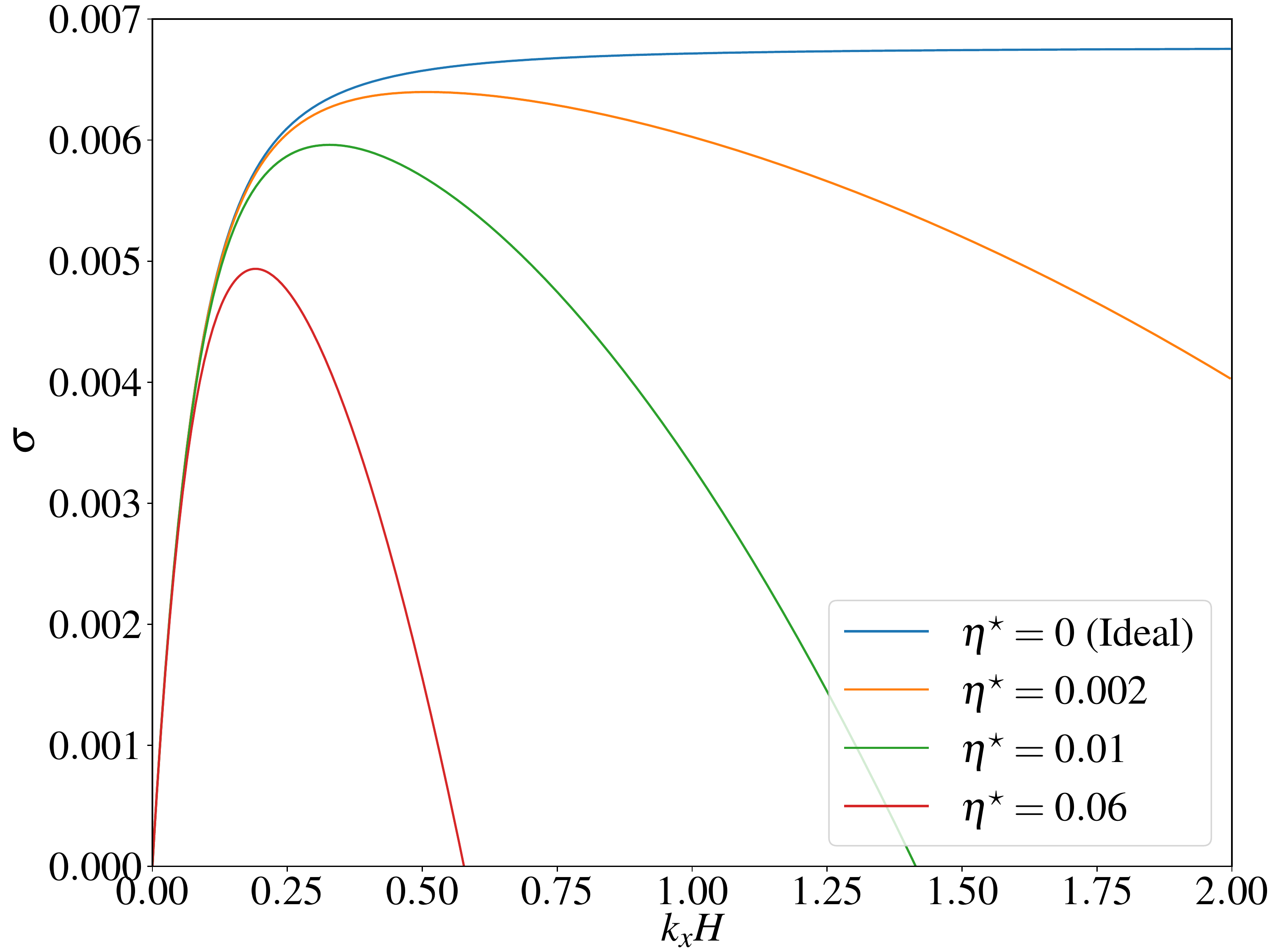}
 \caption{Growth rate of the wind-driven instability as a function of the radial wavenumber for different magnetic diffusivities. Parameters of the model have been chosen in agreement with numerical simulations for  fiducial $\mueq=10^{-3}$:  $\alpha_{\nu}\simeq 0.3$, $p\simeq1$, $q\simeq0.5$, $\zeta_0 \simeq 0.01$.}
\label{fig_growth_rate}
\end{figure}
\subsection{Stability criterion and growth rates}
\label{stability_criterion}
The linear system of equations \ref{mass_eq_lin}, \ref{mx_eq_lin}, \ref{my_eq_lin} and \ref{bz_eq_lin} can be simplified and cast into a (3x3) matrix problem whose determinant is
\begin{equation}
\label{eq_determinant}
D = 
\left|
\begin{array}{ccc} 
\sigma+\zeta_0 & ik_x &  p\zeta_0 \\
\dfrac{\sigma}{2}+\alpha_{\nu_0} & -\dfrac{i}{2k_x} & \alpha_{\nu_0}q\\
\sigma+\eta^\star k_x^2 & 2ik_x& \sigma+\eta^\star k_x^2  \\
\end{array}
\right|
\end{equation}
Solutions of the problem are found by setting $D=0$. It is  straightforward to show that growth rates follow the dispersion relation
\begin{equation}
\label{eq_sigma}
A \sigma^2 + B\sigma + C = 0
\end{equation}
with 
\begin{multline*}
A = 1+ k_x^2 \\
B=   2k_x^2 \alpha_{\nu_0}\left (1+q \right)- \zeta_0 \left (p-1+2pk_x^2\right) +\eta^\star \left[k_x^2+k_x^4\right] \\
C=-   4  \zeta_0 k_x^2 \alpha_{\nu_0}\left (p-q \right)  + \eta^\star k_x^2 \left[\zeta_0(1-p)+2\alpha_{\nu_0}(1-q)k_x^2\right]\\
\end{multline*}
There are actually two different regimes, depending on the strength of the turbulent stress. The first one corresponds to $B<0$, i.e $\alpha_{\nu_0} \ll \zeta_0$ and $k_x^2>(1-p)/(2p)$. In that regime,  solutions are always unstable ($\Re(\sigma)>0$).  The radial flow that concentrates magnetic flux is directly produced by the inertial term $\sigma \hat{v}$, via the geostrophic equilibrium and the conservation of angular momentum. This particular regime, though quite exotic and allowing the instability for $\alpha=0$  is however never encountered in simulations. The second regime corresponds to $B>0$ or  $\alpha_{\nu_0} \gg  \zeta_0$. In that case, the instability occurs if $C<0$, which corresponds in the ideal limit  ($\eta^\star =0$), to 
\begin{equation*}
p>q
\end{equation*}
If $p<1$ and $q<1$, the non-ideal term in $C$ is always positive.  Therefore it contributes to weaken the growth rate. Note that the hypothetical case $q>1$ and $p=\zeta_0=0$  (no winds) can in principle lead to an instability in presence of finite resistivity  but is excluded by the simulations. It corresponds to the usual, but hypothetical "viscous" instability, like imagined by \citet{lightman74}.  

In the ideal limit ($\eta^\star=0$), it is straightforward to show that the optimal growth rate is obtained in the limit $k_x\gg1$.  Under the condition $\alpha_{\nu_0} \gg \zeta_0$, which is verified in most of simulations,  it can be demonstrated that the optimal growth rate, at leading order, is independent of $\alpha_{\nu_0}$ and proportional to $\zeta_0$: 
\begin{equation*}
\sigma \simeq 2\zeta_0 \left( \dfrac{p-q}{1+q}\right)
\end{equation*}

For typical MRI-driven turbulence with $\mueq=10^{-3}$,  realistic values of the parameters are $\alpha_{\nu}\simeq 0.3$, $p\simeq1$, $q\simeq0.5$ and $\zeta_0 \simeq 0.01$.  Using these values, we find in the ideal limit $\sigma = 0.00666$. To illustrate the effect of magnetic diffusivity, we plot in Fig.~\ref{fig_growth_rate} the growth rate of the instability $\sigma$ as a function of $k_x$ for three different resistivities ($\eta^\star\simeq 0.002,0.01,0.06$).  These values are here purely ad-hoc and just to illustrate the dependence of $\sigma(k_x)$ on $\eta^\star$.  Realistic values,  directly inferred from 2D and 3D simulations,  are used in Section \ref{sec_simulations} and \ref{sec_simulations3D}.  In all cases the maximum growth rate is obtained for $k_x H\lesssim 0.5 H$ which correspond to rings of size $r_a\gtrsim 10\, H$. The instability  occurs on long timescale $1/\sigma$ larger than $150\, \Omega^{-1}$ (20 orbits).

\subsection{Eigenmodes}
The linearised system admits solutions of the form: 
\begin{equation}
\label{eq_eignemode}
\left(\begin{array}{c}
\hat{u} \\
\hat{v} \\
\hat{b}_z
\end{array} \right)
= \left(\begin{array}{c}
-i k_x E \\
 i {k_x}/{2}  \\
 - {k_x^2 E}/{(\sigma+\eta^\star k_x^2)} \\
\end{array} \right)  \, \hat{\Sigma}
\end{equation}
with $E= \left[(\sigma/2+\alpha_{\nu_0})p\zeta_0 - \alpha_{\nu_0}q (\zeta_0+\sigma)\right]/(p\zeta_0/2 +k_x^2 \alpha_{\nu_0}q)$ a small number compared to 1. In the turbulent regime with $\alpha_\nu \gg \zeta$, and if $p>q$, it is straightforward to show that $E>0$ . The radial  velocity perturbation is out of phase (with an angle of $\pi/2$)  with respect to the density maximum and anti-correlated with the zonal flow $\hat{v}$. The vertical field and the magnetization are anti-correlated with the density maximum. This configuration is exactly that depicted in Section \ref{sec_phenomenlogy}. 

\section{2D axisymmetric simulations}
\label{sec_simulations}
To test our instability model, we need to confront the theoretical predictions of Section \ref{sec_theory} with MHD numerical simulations exhibiting rings structures.  As a preliminary check, we performed in appendix \ref{appendixC} a stability analysis around an initial laminar wind equilibrium, using 2D shearing box simulations. We found that axisymmetric perturbations undergo clean exponential growth with rates and eigenfunctions compatible with our theory.  In this section, we explore the case of a turbulent disc, by using 2D simulations initialized with a net vertical field and random noise. In particular, we check whether the numerical growth rate and spacing of axisymmetric modes, as well as their physical behaviour,  are compatible with the linear theory.  
\subsection{Numerical setup}
Shearing-box simulations are run with the PLUTO code \citep{mignone07}, a finite-volume method with a Godunov scheme that integrates the compressible MHD equations in their conservative form. The fluxes are computed with the HLLD Riemann solver for runs without ambipolar diffusion and with HLL otherwise. The gas is isothermal and inviscid (no explicit viscosity). Boundary conditions are shear-periodic in $x$ and periodic in $y$. 
In the vertical direction, we use standard outflow boundary conditions for the velocity field and impose hydrostatic balance in the ghost cells for the density. In this way, we reduce significantly the excitation of artificial waves near the boundary. For the magnetic field, we use the "vertical field" or open boundary  conditions with $B_x=0$ and $B_y=0$ at $z=\pm L_z/2$. Because the instability we are seeking occurs on a long timescale,  it is important to maintain a constant 
disc surface density $\Sigma$. Otherwise, the wind would empty the disc before the instability reaches any saturated state. For that purpose, we regularly inject mass near the midplane at each numerical time step. The source term in the mass conservation equation is similar to the one prescribed by \citet{lesur13},
\begin{equation}
\dot{\sigma}_i=\dfrac{2\dot{\rho_i}(t)}{\sqrt{2 \pi} z_i}\exp{\left(-\dfrac{z^2}{2z_i^2}\right)}, 
\label{injection}
\end{equation}
where $\dot{\rho_i}(t)$ is the mass injection rate computed at each time step to maintain the total mass constant in the box, and $z_i$ is a free parameter that corresponds to the altitude below which mass is replenished. Note that $\sigma_i$ is uniformly distributed in  $x$ and $y$ and thus independent of local density variations in the box.  

For most of the simulations, we chose a large  horizontal box size $L_x=L_y=20\, H$ to be able to capture the largest rings.  In $z$, the box spans -4 $H$ to 4 $H$. We adopt a resolution of 256 points in the horizontal directions and 128 points in the vertical direction. In the ideal regime, such resolution is insufficient to resolve the small-scale MRI turbulence properly, but enough to obtain the right axisymmetric dynamics. We checked that doubling resolution ($512 \times 256$) does not change the results in terms of growth rate and spacing. 

Finally, for simulations with ambipolar diffusion, we use exactly the same setup as \citet{riols18} where the ambipolar Elsasser number Am is 1 in the midplane and increases abruptly above a certain height corresponding to the FUV ionization layer (see Section 2.4 of the paper for more detail about the prescription and its physical motivation). 

\begin{figure}
\centering
\includegraphics[width=\columnwidth]{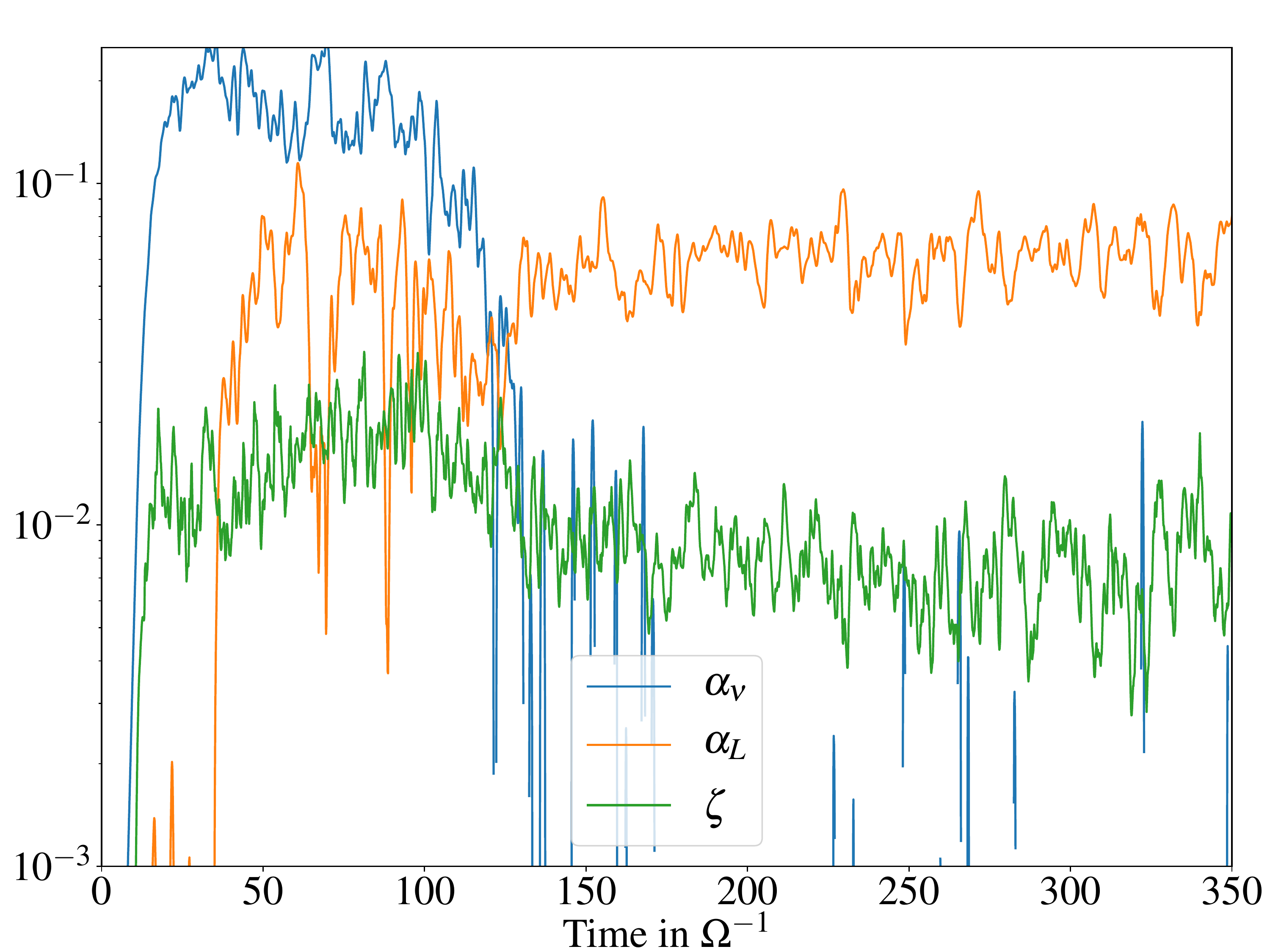}
 \caption{Evolution of the transport coefficient $\alpha_\nu$, $\alpha_L$  and the wind mass loss efficiency $\zeta$ in the 2D simulation without explicit diffusion ($ \mueq=10^{-3}$) }
\label{fig_nonlinear}
\end{figure}
\begin{figure*}
\centering
\includegraphics[width=\textwidth]{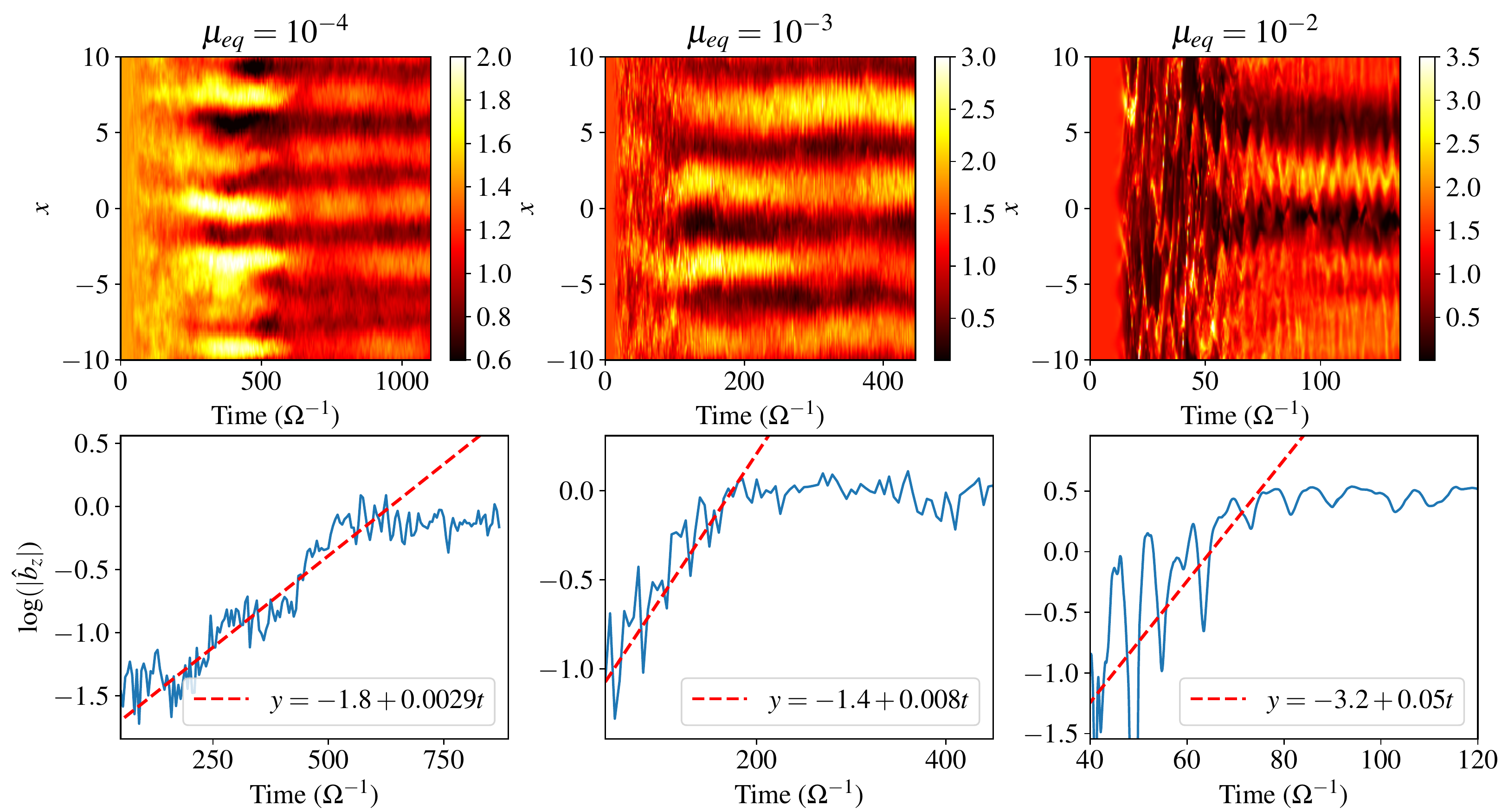}
 \caption{Top: space-time diagram showing the column density  as a function of time and $x$, in 2D axisymmetric turbulent simulations, for different magnetization (left to right, $\mueq=10^{-4},10^{-3}$ and $10^{-2}$). The density is integrated within $z\pm 1.5 H$ and runs are computed in the ideal limit (without explicit magnetic diffusion). Bottom panels show for each run, the time-evolution of $\hat{b}_z$, the normalized projection of $B_z$ onto the prominent axisymmetric Fourier mode (with largest amplitude). This mode corresponds  respectively to $k_x=6$, $4$ and $2 k_{x_0}$ for $\mueq=10^{-4},10^{-3}$ and $10^{-2}$, with $k_{x_0}=2 \pi/L_x$ the fundamental box radial wavenumber.}
\label{fig_bzhat}
\end{figure*}
\begin{figure}
\centering
\includegraphics[width=\columnwidth]{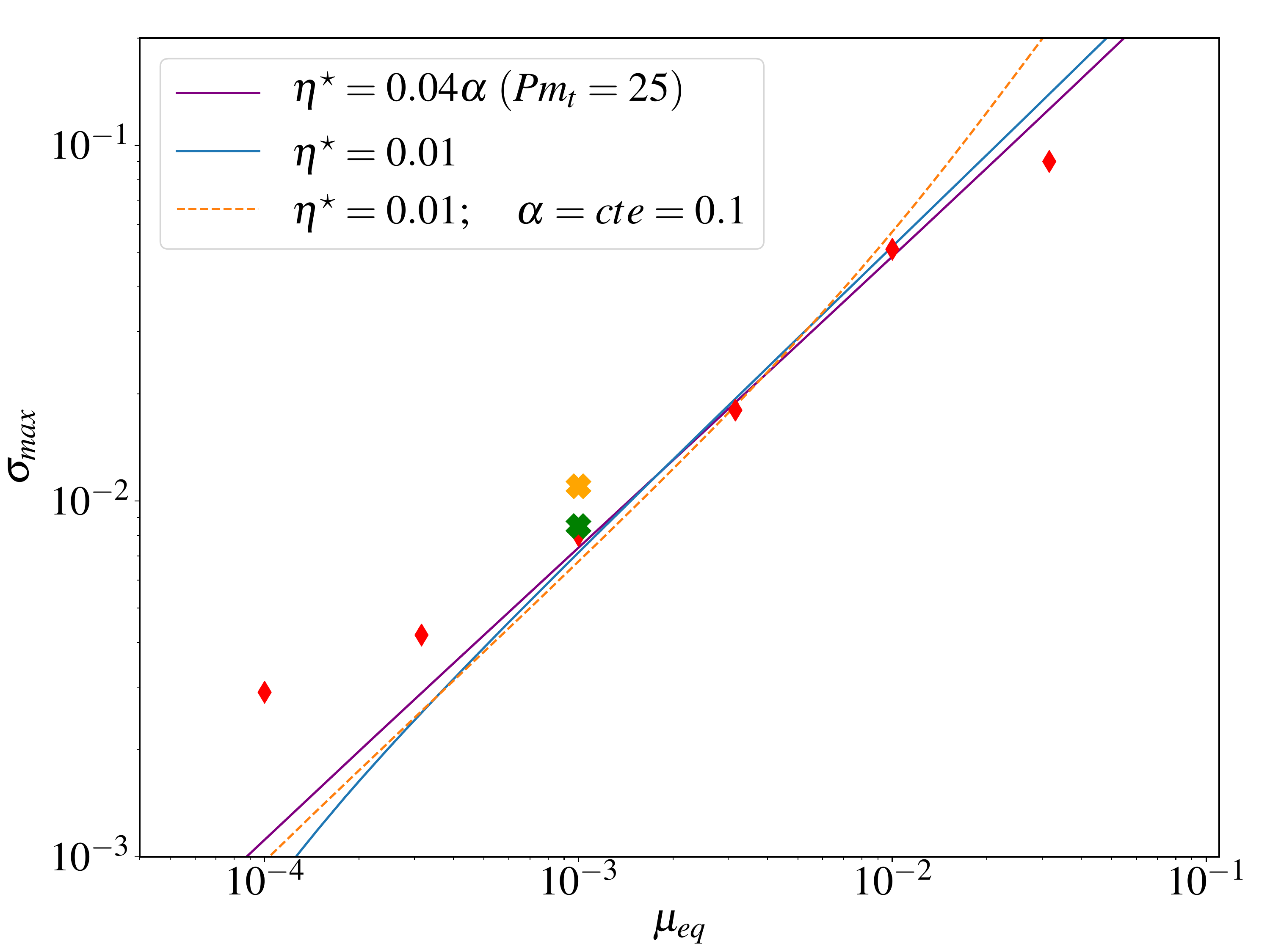}
 \caption{Optimal growth rate of the instability as a function of the magnetization $\mu$. The red diamond markers are the growth rates of the most unstable Fourier mode measured in 2D ideal simulations. The orange and green crosses are those measured in resistive and ambipolar runs.  The purple plain line correspond to the model prediction, using the scaling relations (\ref{eq_scaling1}) and (\ref{eq_scaling2}).  The blue and yellow/dashed lines use the same model but with constant $\eta^\star=0.01$ in both cases and constant $\alpha_\nu=0.1$ for the last case.}
\label{fig_growth_rate_Bz}
\end{figure}
\begin{figure}
\centering
\includegraphics[width=\columnwidth]{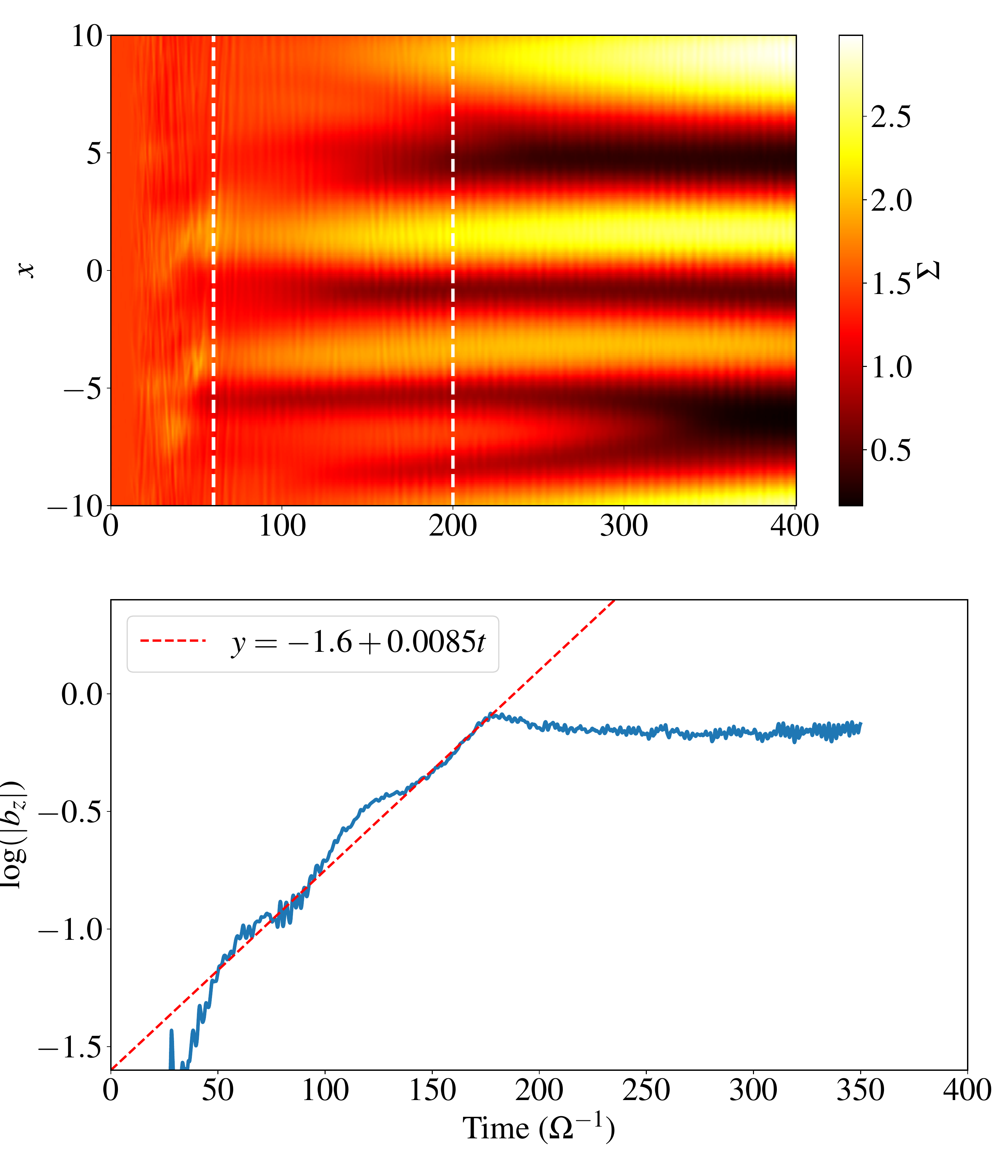}
 \caption{Top panel:  spacetime diagrams showing the surface density $\Sigma(x,t)$  in the 2D ambipolar simulation ($\text{Am}_{mid}=1$,  $\mueq=10^{-3}$ and  $z_d=2 H$). The two vertical dashed lines delimit the "linear" phase, during which zonal modes, in Fourier space,  grow exponentially. The bottom panel shows the evolution of the $B_z$ component  projected onto the $k_x=4 k_{x_0}$ mode.}
\label{fig_2Dambipolar}
\end{figure}
\begin{figure}
\centering
\includegraphics[width=\columnwidth]{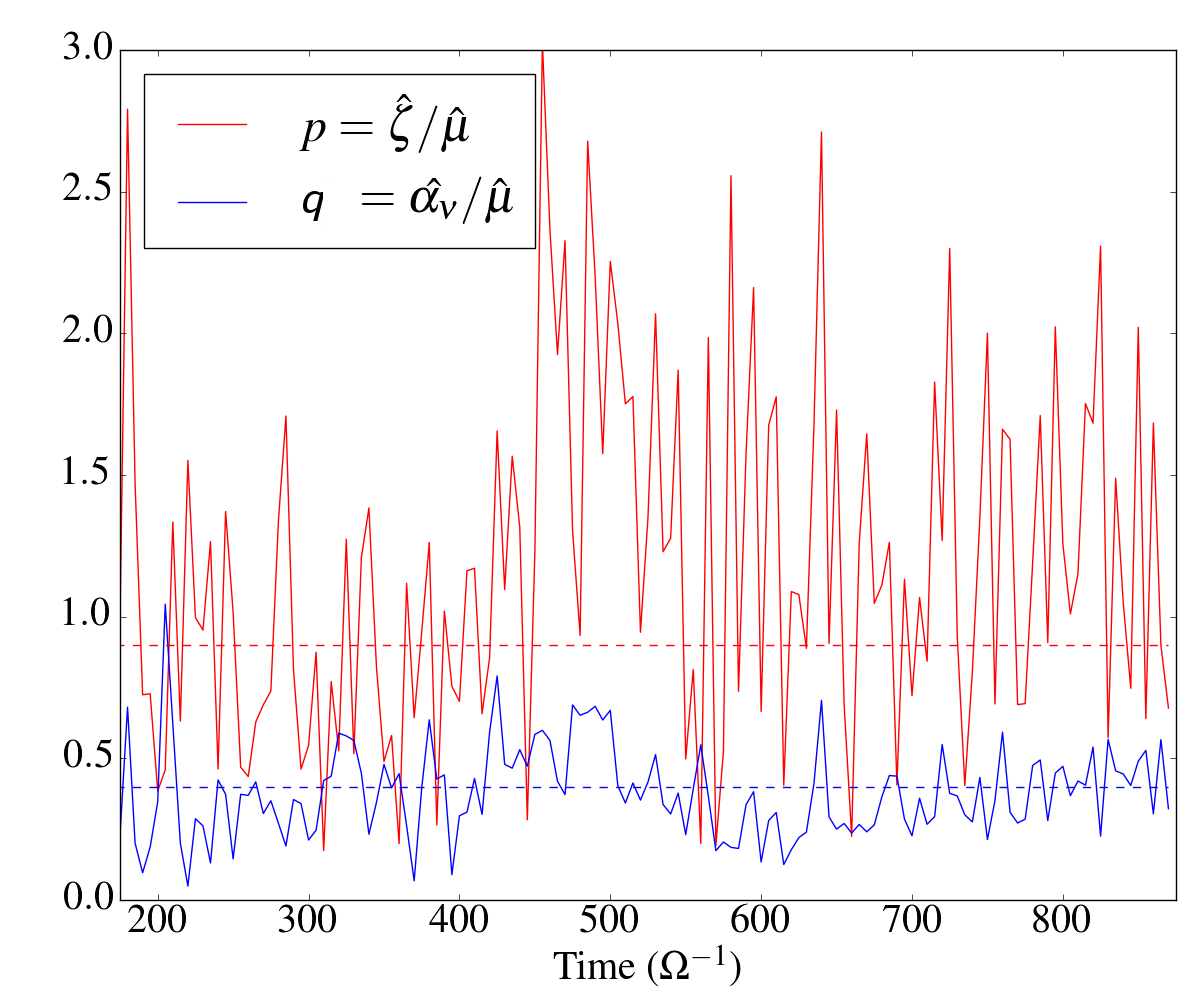}
 \caption{Coefficients $p$ and $q$ measured from the 2D ideal simulation with $\mueq=10^{-4}$. The red  (respectively blue) line shows the ratio between the perturbation of $\zeta$ (respectively $\alpha$) and the perturbation associated with magnetization, as a function of time.  Each quantity ($\zeta$, $\alpha$ and $\mu$) is  projected onto the mode  $k_x=6 k_{x_0}$, averaged within $z \pm 2 H$ and normalized to its mean.}
\label{fig_scaling2d}
\end{figure}
\subsection{Simulations in the ideal limit and numerical growth rates}
\label{2D_ideal}
We first run a series of 2D axisymmetric simulations in the ideal limit (without explicit resistivity or ambipolar diffusion) by varying the vertical magnetization $\mueq$ from $10^{-5}$ to $0.03$.  All runs are initialized with an hydrostatic equilibrium in density and a weak random noise in velocity.
Figure \ref{fig_nonlinear} shows the evolution of the turbulent and laminar transport coefficients $\alpha_{\nu}$, $\alpha_{L}$ and the wind loss efficiency $\zeta$ for the case $\mueq=10^{-3}$. To complement the analysis, the evolution of column density $\Sigma(x)$  is illustrated in   Fig.~\ref{fig_bzhat} (top panels) for three different magnetization $\mueq=10^{-4}$, $10^{-3}$ and $10^{-2}$. Each quantity is integrated between -1.5 and 1.5 $H$.  In all cases, we identified three distinct phases associated with : a) the development of  vigorous MHD turbulence and the launching of a wind during the first $\sim 50-150 \,\Omega^{-1}$, b) the growth of zonal flows and density rings, and c) the saturation of zonal flows which is accompanied by a severe drop in the turbulent stress and a modest drop in the wind loss efficiency $\zeta$ (see Fig.~\ref{fig_nonlinear}).  Note that during the initial turbulent phase, $\alpha_{\nu} \gg \alpha_{L}$, while after the saturation of sub-structures,  the radial stress is predominantly laminar $\alpha_{\nu} \ll\alpha_{L}$. Figure \ref{fig_bzhat} shows that the timescale associated with rings formation seems to decrease with $\mueq$ while their spacing and strength seems to increase with $\mueq$. 

To understand whether rings form via a linear instability or a more complicated non-linear process, we explore the dynamics of axisymmetric modes in Fourier space. The procedure is simple: at each time-step, we perform a 1D FFT (along the $x$ direction) of the vertically averaged quantities.   We then project each field (or any physical quantity) onto the dominant axisymmetric mode that grows during the simulation \footnote{This is defined as the mode with maximum amplitude, which corresponds to the $k_x=4 k_{x_0} = 8\pi/L_x$  mode at $\mueq=10^{-3}$. At low magnetisation ($\mu=10^{-4}$) however zonal flow cannot be properly reduced to a single mode.  The peak of the spectrum oscillate between the $k_{x}=5$ and $k_{x}=6 k_{x_0}$ components. }. In this way, we keep the interesting dynamics related to the prominent ring and filter out part of the dynamics associated with the turbulent flow.  For a given field $\varphi$, we note $\hat{\varphi}$ such projection, normalized with respect to its mean (the $k_x=0$ mode in Fourier space).  

Figure \ref{fig_bzhat} (bottom panels) shows the evolution of $\hat{b}_z$ for three different magnetizations. In all cases, the prominent ring  mode starts growing quasi-exponentially, indicating that a linear instability is at work.  Note that the projection procedure is necessary to obtain a clean exponential growth at the beginning of simulations.  After a few tens of orbits, the growth stops and $\hat{b}_z$ saturates around 1. This is an indication that a non-linear regime is reached. 

During the linear phase,  we measure the growth rates associated with the prominent mode and report them in Fig.~\ref{fig_growth_rate_Bz} for different $\mueq$.  For $\mueq> 10^{-4}$, growth rates increase with the magnetization and vary from 0.002 to 0.05 $\Omega$; the dependence of $\sigma$ with $\mueq$ follows a power law with index $~0.7-0.8$. We checked that $\sigma$ is not changed when doubling the resolution of simulations. It is worth noting that such growth rates are $\sim 10-20$ smaller than those characterizing  the MRI at a similar scale. 

\subsection{Non-ideal case}
\label{2D_nonideal}
We first investigate the effect of ohmic resistivity  on the growth of axisymmetric structures. For that, we run a 2D simulation with $\mueq=10^{-3}$ and explicit  $\eta /(\Omega H^2)= 0.01$. This corresponds to a magnetic Reynolds number 
$\text{Rm}=\Omega H^2/\eta=100$. 
During the first tens of orbits, a turbulent state develops but with a much weaker strength and transport than in the ideal case  (for comparison,  $\alpha \simeq 0.05$). Such a result is expected since the MRI is quenched by the resistivity. However the wind mass loss efficiency $\zeta$   is of the same order of magnitude. Three rings, associated with a mode $k_x=3k_{x_0}$ develop in the box and grow at a rate $\sigma \simeq  0.011  \Omega $. The number of rings, their properties, and the growth rate are actually comparable to those obtained in the ideal case. 

We then study the effect of ambipolar diffusion by running a simulation with the same $\mueq=10^{-3}$ and $\text{Am}=1$ in the midplane. Figure \ref{fig_2Dambipolar} shows the evolution of surface density and vertically-averaged $B_z$. Again tree or four structures seem to emerge from the initial turbulent phase. The growth rate associated with the $k_x=4 k_{x_0}$ mode in Fourier space is $\sigma\simeq 0.0085$, very similar to the value obtained in the ideal limit. Thus, ambipolar diffusion does not seem to alter the instability mechanism, although it weakens or even suppresses the initial turbulent state. 

Note that unlike the ideal case,  the turbulent transport is either comparable to or smaller than the laminar transport $\alpha_L$. To go further in the analysis and check that the instability identified numerically is of same nature as that described in Section \ref{sec_theory}, we inspected in the ambipolar run the different flux and source terms in equations \ref{mass_eq}, \ref{mx_eq},  \ref{my_eq}  and \ref{bz_eq}.  For ease of reading, the analysis is done in Appendix \ref{appendixD}. We show in particular that the ambipolar term $ \eta_A \mathbf{J} \times (\mathbf{B} \times \mathbf{B})/B^2 $ merely acts as a diffusion on the $B_z$ field. 

\subsection{Parameters $p$ and $q$ and confrontation with the model}
\label{scaling_law}
In this section, we  investigate whether the dependence of numerical growth rates $\sigma (\mueq)$  can be predicted by the simple model exposed in Section \ref{sec_theory}. We focus particularly on the ideal simulations, for which $\alpha_{L_0}\ll \alpha_{\nu_0}$ during the linear phase. The linear theory can actually be generalized to quasi-laminar discs with $\alpha_{L_0}\gtrsim\alpha_{\nu_0}$ (see Appendix \ref{appendixB}), typically those obtained in non-ideal simulations.  The instability in this regime is conceptually not different and the theoretical growth rates are comparable to those obtained with a pure turbulent stress.

The model of Section \ref{sec_theory} depends on three main parameters  $p=d\log(\alpha_\nu)/d\log(\mu)$,  $q=d\log(\zeta)/d\log(\mu)$ and the diffusivity $\eta_\star$. 
To evaluate these coefficients, it is suitable to measure them directly from numerical simulations.  A naive way is to run a series of turbulent simulations varying $\mu$ and measuring $\alpha_\nu$, $\zeta$ and $\eta_\star$. Though simple, this method is complicated to accomplish in practise.  Indeed, as suggested by Fig.~\ref{fig_nonlinear}, the quasi-steady turbulent state obtained in the early stage of simulations is short ($< 100 \Omega^{-1}$) and rapidly affected by the zonal structures.  This turns to be particularly critical at large $\mueq$, for which measures of transport coefficient and mass loss rate are not statistically meaningful and can be highly inaccurate.\\

A possible way to circumvent this issue is to infer directly the scaling relations from a thorough inspection of the large-scale ring perturbations themselves.  The idea is to compute the perturbed coefficients  $\hat{\alpha}_\nu$ and $\hat{\zeta}$  associated with the dominant axisymmetric mode and see how it correlates in time with the perturbed  magnetization $\hat{\mu}$.  In addition to giving $p$ and $q$, it provides a great opportunity to check the linear relations conjectured in Section \ref{linearisation}, which are fundamental requisites of  the instability.  \\

Figure \ref{fig_scaling2d} shows the two ratios $p=\hat{\zeta}/\hat{\mu}$ and $q=\hat{\alpha}_\nu/\hat{\mu}$ for the case $\mueq=10^{-4}$. Here $\hat{\alpha}_\nu$,  $\hat{\zeta}$ and  $\hat{\mu}$ are the vertically averaged perturbations of transport, mass loss efficiency and magnetization projected onto the Fourier mode $k_x=6\times 2\pi/L_x$ and normalized with respect to the mean ($k_x=0$  mode). Note that the latter is averaged  in time during the growth phase.   Although these ratios are highly fluctuating (this is particularly true for $p$),  $\hat{\zeta}$ and  $\hat{\alpha}$ seem linearly correlated to the magnetization $\hat{\mu}$, at least statistically.  Most importantly, we check that  $p>q$. An interesting but unexpected result is that the ratios seem to keep a similar value in the saturation regime ($t \gtrsim 500 \Omega^{-1}$).  By averaging in time, we find $q \simeq 0.4$ and $p\simeq 0.9$ in agreement with the scaling relations obtained in 3D fully turbulent simulations  \citep{scepi18}.   We did the same calculation for  $\mueq=10^{-3}$ and found $p\simeq 0.8$ and $q\simeq 0.55$.  

In sum, we adopt in our model the following scaling laws: 
\begin{equation}
\label{eq_scaling1}
\alpha_\nu=   4\, \mu^{0.45}  \quad \text{and} \quad \\
\zeta =  4.5 \, \mu^{0.8}
\end{equation}
with the constants calibrated to fit with the simulation data at intermediate $\mu= 10^{-3}$ (values are taken from Fig.~\ref{fig_nonlinear}). Simulations in this regime generally provide a more accurate measure of $\alpha_{\nu_0}$ and $\zeta_0$ since the turbulent phase settles longer. The dependence of $\alpha$ is very close to that obtained  in past 3D simulations \citep[see Eq.~20 of][]{salvesen16}. 

To estimate the turbulent magnetic diffusivity, we use a slightly different method. Instead of projecting the flow into the Fourier space, we calculate directly in real space  the averaged term  $\langle \mathcal{E}_y\rangle - \langle \underline{v_z} \rangle \langle B_x \rangle +  \langle\underline{ v_x }\rangle  \langle B_z \rangle$ during the linear growth of the mode.  We checked that it correlates quite well in space with ${\partial \langle B_z \rangle }/{\partial x}$ and indeed acts to diffuse the large scale structure.   The ratio between the two terms is $\eta_t /(\Omega H^2) \approx 0.003$ for $\mueq=10^{-4}$ and $\eta_t \approx 0.008$ for $\mueq=10^{-3}$. This corresponds to rather low turbulent diffusivities,  compared to values given by  \citet{lesur09b} in the unstratified 3D case. The turbulent Prandlt number $\text{Pm}= \alpha_\nu/{\eta_t}$ is of order 25 and we checked that it depends little on resolution. Therefore, we consider:
\begin{equation}
\label{eq_scaling2}
{\eta_t}= 0.04\, \alpha_\nu \\
\end{equation}
These scaling laws are then used as inputs of our model (Section \ref{stability_criterion}), and allow us to compute the optimal growth rate as a function of $\mu$ (using Eq.~\ref{eq_sigma}).  The result  is superimposed  in Fig.~\ref{fig_growth_rate_Bz} (purple curve). The model fits quite well with the numerical values obtained in Section \ref{2D_ideal}, and in particular reproduces the slope of $\sigma(\mu)$ for intermediate  $\mu$.  
We stress that such a result does not rely exclusively on the linear correlations measured in Fig.~\ref{fig_scaling2d}, since growth rates depend also on the absolute values of $\zeta$. To show that the model is robust, we also plot the theoretical growth rates obtained for a constant magnetic diffusivity (blue curve) and constant $\alpha$ and $\eta^\star$ (yellow dashed curve). In both cases, there are few differences with the prescription ${\eta_t}= 0.04\, \alpha_\nu$.   This is not surprising since maximum growth rates depend little on $\alpha$ and $\eta^\star$  (for the range of diffusivities considered here,  see Fig.~\ref{fig_growth_rate}). 

We finally comment on the extreme cases, corresponding to the lowest and highest magnetizations in Fig.~\ref{fig_growth_rate_Bz}.  In these regimes, the numerical values slightly deviate from the model.  For $\mu \gtrsim  10^{-1.5}$, measures are perhaps inaccurate since growth rates associated with large-scale modes become comparable to those associated with the MRI phase.  For $\mu\lesssim 10^{-4}$, the small discrepancy is probably due to a sudden drop in $\alpha$ and $\eta_t$. The fact that the turbulence dies out at low $\mu$ is expected since the 2D box does not sustain a dynamo in the limit $\mu \rightarrow 0$. As turbulent dissipation is weakened, zonal flows are slightly enhanced. This effect, however, remains marginal and only hold for $ 10^{-5} \lesssim\mu \lesssim 10^{-4}$.  Below $\mu=10^{-5}$,  the vertical flux of mass associated with the wind drops abruptly, and we checked  numerically that zonal flows vanish.

\begin{figure}
\centering
\includegraphics[width=\columnwidth]{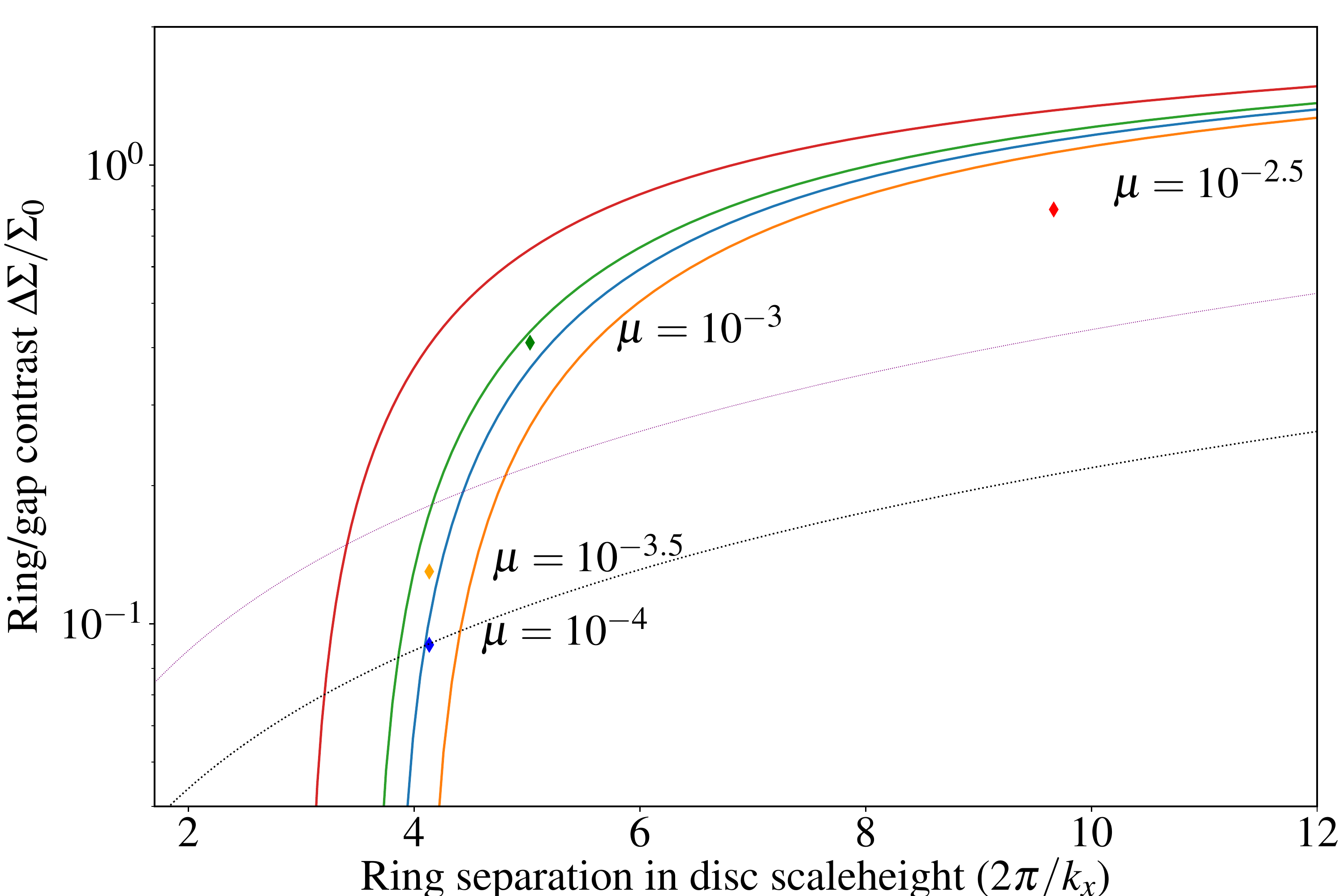}
 \caption{Ring/gap contrast versus radial separation. The plain curves are estimated from the theory (Eq.~\ref{eq_contrast})  while the diamond markers are points measured from simulations. The blue,  orange, green and red colors corresponds respectively to $\mueq=10^{-4}, 10^{-3.5}, 10^{-3}$ and $10^{-2.5}$. Dotted lines delimit the region above which dust can be concentrated within the rings, assuming $H/R=0.05$ (black) and $H/R=0.1$ (purple).  See Section \ref{dust_concentration} for more details.}
\label{fig_contrast}
\end{figure}
\subsection{Nonlinear saturation and ring/gap contrast}
\label{contrast_ring}
As suggested  by Fig.~\ref{fig_bzhat}, the ring instability saturates and enters a non-linear regime, once the azimuthal structures reach a significant amplitude.   During this phase, the structures stop growing but prevail in the flow and remain stable for hundreds of orbits.  Their non-linear saturation leads to a drop in the mean stress (Fig.~\ref{fig_nonlinear}) and the production of a strong mean azimuthal field $B_y$, altering significantly the initial hydrostatic density profile.  
But how does this saturation occur and what determines the final amplitude (or density contrast) of the rings?  By examining the density contrast in the case $\mu=10^{-4}$ or $\mu=10^{-3}$ (upper panels of Fig.~\ref{fig_bzhat}), we find that the gaps in the nonlinear regime still  contain a large amount of gas.  For instance,  the surface density perturbation associated with the prominent Fourier mode, settles toward  $\hat{\Sigma}\sim 0.09$ for $\mueq=10^{-4}$ and $\hat{\Sigma}\sim 0.42$ for $\mueq=10^{-3}$. Therefore, saturation does not occur because the material in the gaps has been emptied. The lower panels of Fig.~\ref{fig_bzhat} show instead that the instability saturates when $\hat{b}_z \sim 1$  (i.e perturbation of ${B}_z \sim B_{z_0}$). In other words, the instability stops when there is no more vertical flux to drag in.   The density contrast between the gaps and the rings in the nonlinear regime can be then estimated by setting $\hat{b}_z \sim 1$ in Eq.~\ref{eq_eignemode}. We obtain in the ideal limit:
\begin{equation}
\label{eq_contrast}
\Delta\Sigma/\Sigma_0 = \vert \hat{\Sigma} \vert\simeq \dfrac{\sigma}{k_x^2 E(\sigma,k_x)}  \simeq \left[\dfrac{\zeta}{\sigma} \left(\dfrac{p}{q} -1\right ) -1\right]^{-1}
\end{equation}
where the last equality is obtained in the limit $k_x^2 \alpha_{\nu_0}/\zeta_0 \gg 1$. Figure \ref{fig_contrast} shows the theoretical density contrast for different magnetisations (plain lines) as a function of $k_x$. This is calculated using the same parameters  and scaling relations as in Section \ref{scaling_law}. The result is that $\Delta\Sigma/\Sigma_0$  depends mainly on the radial separation, which is a function of the magnetization.   We report on the same figure the density contrast of leading axisymmetric modes (respectively $k_x=2, 4, 5 ,5 k_{x_0}$ measured in simulations for $\mueq=10^{-4}$, $10^{-3.5}$, $10^{-3}$ and $10^{-2.5}$. Numerically,  there is a close relationship between the density contrast and the rings separation, which appears consistent with  the theoretical calculations.
\subsection{Ring separation}
So far, we simply measured the ring separation as the most prominent $k_x$ in simulations, but can we predict this quantity from the linear theory? For large magnetization ($\mueq=10^{-2}$ and $10^{-2.5}$), this corresponds roughly to the radial scale that maximizes the theoretical growth rate.  However, for smaller magnetization, this is no more the case.  For instance, the prominent modes in simulation with $\mueq=10^{-4}$ corresponds to $k_x=5 k_{x_0}=1.5H^{-1}$ and  $6k_{x_0}=1.8 \,H^{-1}$, while the linear theory gives a maximum growth rate at $k_x\simeq 0.45\, H^{-1}$. Actually, this is not in contradiction with the theory since $\sigma$ appears to be quite flat with $k_x$ in this regime. The reason is that turbulent dissipation is low ($\text{Pm}_t=25$) and permits small scales axisymmetric modes to be amplified with significant growth rates. Modes then compete with each other and it becomes extremely difficult to predict the spacing from linear theory. The initial amplitude will be probably a decisive factor  in the selection of the dominant mode(s). Such amplitude depends on the spectrum of the initial turbulence phase, which is a priori difficult to predict. We found (not shown here) that the spectrum of $B_z$  associated with the initial turbulence peaks around $k_x=6k_{x_0}$ and beyond for $\mueq=10^{-4}$, while it peaks at much lower $k_x\lesssim 0.3 $ in the case $\mueq=10^{-2}$. For $\mueq=10^{-4}$, there is roughly a factor 25 in initial energy between the  $k_x=6k_{x_0}$ mode and the large-scale mode which, in principles, should grow fastest  ($k_x\simeq 0.45\, H^{-1}$). Therefore, it is likely that the ring spacing in our simulations is partly ruled by the initial turbulent conditions. Whether the spectrum of this turbulence is universal or depends on the historic and detailed physics of the disc remains an open question.  Note however that all simulations, even in 3D (see Section \ref{sec_simulations3D}),  indicate a similar trend that the ring spacing increases with magnetisation. We finally emphasize that it does not depend too much on horizontal box size (see a comparison with small box \cite{riols18}) or resolution.  

\subsection{Criterion for dust concentration}
\label{dust_concentration}
With the result of Section \ref{contrast_ring}, we can infer a minimum magnetization $\mu$ for which dust can accumulate into the gaps. A simple criterion is that the local pressure gradient in the disc have to be positive:
\begin{equation}
\dfrac{1}{P_0} \left(\dfrac{\partial P}{\partial R}\right) = \dfrac{1}{P_0} \left( \dfrac{\partial P_0}{\partial R}\right)+ k_R \dfrac{ \Delta \Sigma}{\Sigma_0}> 0 
\end{equation}
where $P_0$ is the mean equilibrium pressure profile and $k_R$ the local radial wavenumber of the ring perturbation.  A typical disc model, which matches disc observations \citep{andrews09} has
\begin{equation}
P_0 \propto R^{-11/4}   \quad   \text{and then}  \quad   \dfrac{1}{P_0} \left( \dfrac{\partial P_0}{\partial R}\right) = -\dfrac{11}{4R} 
\end{equation}
The criterion for dust accumulation becomes: 
\begin{equation}
\left(\dfrac{ \Delta \Sigma}{\Sigma_0} \right) \gtrsim  \dfrac{11}{4} \left( \dfrac{H}{R} \right) \frac{1}{ k_R H}
\end{equation}
The black and purple dotted lines in Fig.~\ref{fig_contrast} delimit the region above which a radial concentration of dust is possible, respectively for $H/R=0.05$ and $H/R=0.1$. We predict that magnetizations greater than $10^{-4}$ or $10^{-3.5}$ will lead to the formation of dusty rings.  A more accurate estimation would require to take into account the radial diffusion of particles by the turbulence, which depends on the grain size and the nature of non-ideal effects,  but this is beyond the scope of this paper.  
\section{3D simulations}
We extend our analysis  to 3D simulations and show that ring structures exhibit properties that are again compatible with the instability model of Section \ref{sec_theory}. 
\label{sec_simulations3D}
\begin{figure*}
\centering
\includegraphics[width=\textwidth]{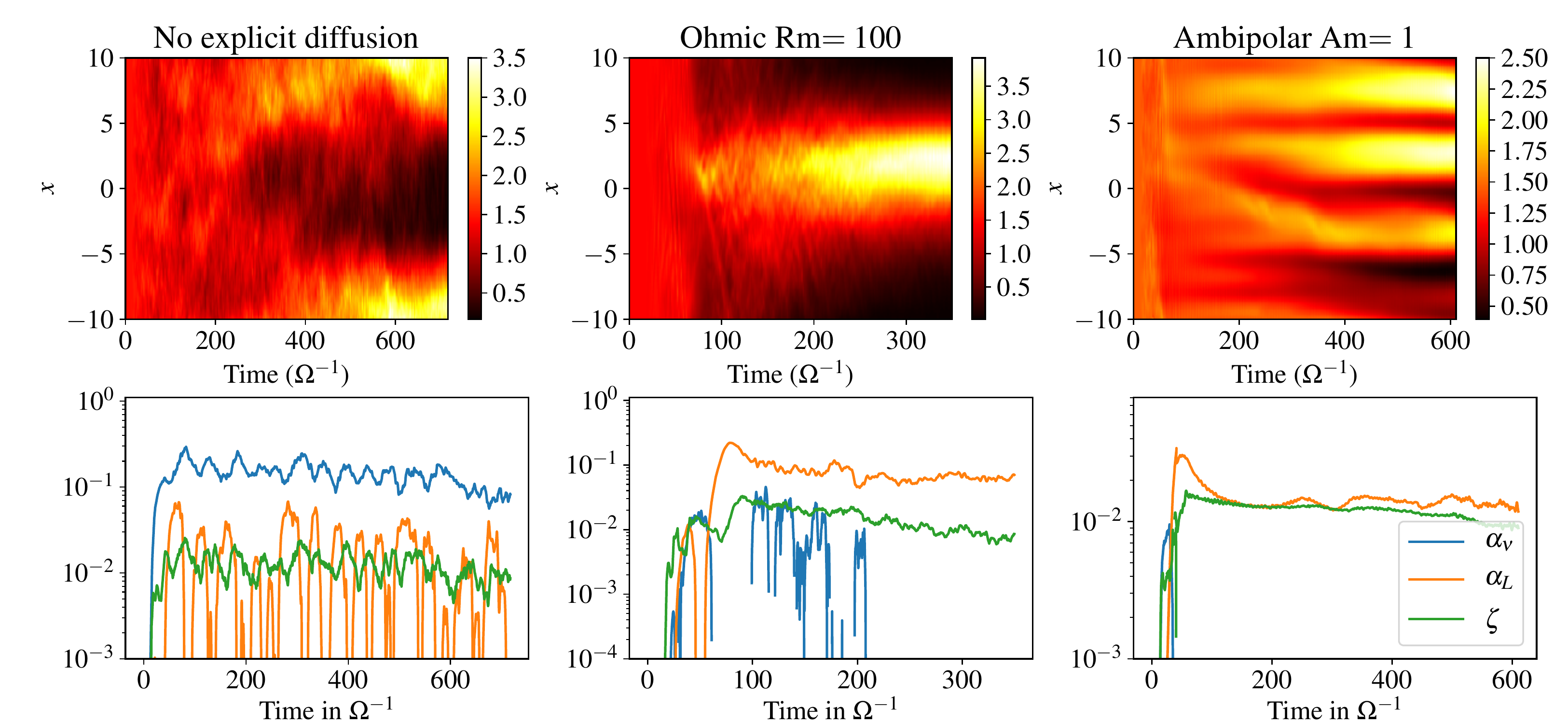}
 \caption{Top: space-time diagram showing the column density  as a function of time and $x$, in 3D turbulent simulations with $\mueq=10^{-3}$. Left panel:  no explicit diffusion; center panel : ohmic diffusion ($\text{Rm}=\Omega H^2/\eta=100$); right panel: ambipolar diffusion ($\text{Am}_{mid}=1$). The density is integrated within $z\pm 1.5 H$.  Bottom panels shows for each run, the time-evolution of turbulent $\alpha_\nu$, laminar $\alpha_L$ and $\zeta$ the mass loss efficiency.}
\label{fig_3d}
\end{figure*}
\begin{figure}
\centering
\includegraphics[width=\columnwidth]{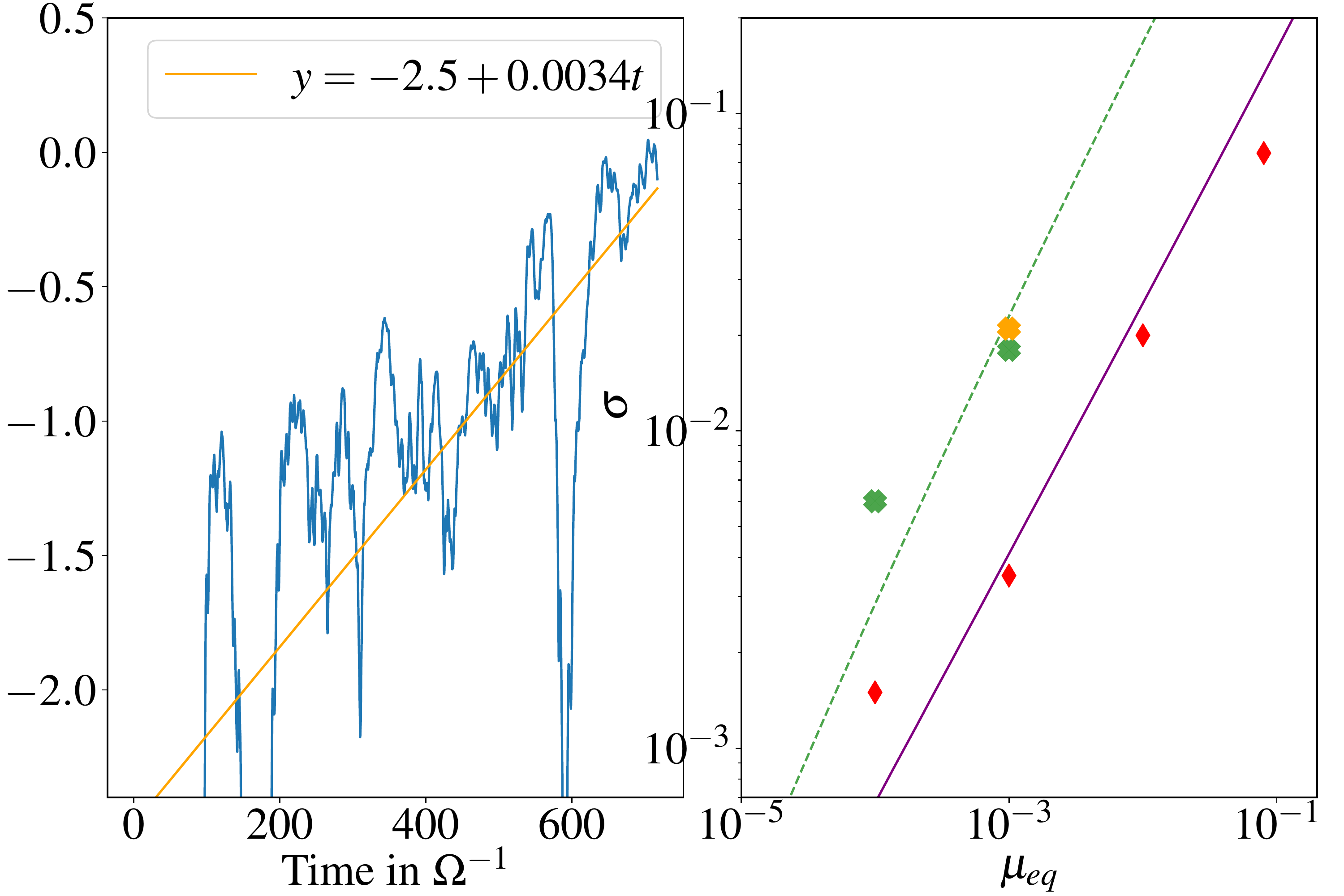}
 \caption{Left: time-evolution of $\hat{b}_z$, the normalized projection of $B_z$ onto the prominent axisymmetric Fourier mode ($k_x=2 \pi/L_x$) in the ideal simulation with $\mueq=10^{-3}$. Right: growth rate of the linear instability as a function of the magnetization $\mu$. The red diamond markers are measures from 3D ideal simulations. The green and yellow crosses are those measured in resistive and ambipolar runs.  The plain lines correspond to model predictions, using similar scaling relations as (\ref{eq_scaling1}), but with $\eta_t=0.5 \alpha_\nu$ in the ideal case.}
\label{fig_3d2}
\end{figure}
\subsection{Dependence on diffusive processes}
First, we perform three different simulations  for $\mueq=10^{-3}$, respectively with no explicit diffusion, ohmic resistivity $\eta=0.01$ and ambipolar diffusion $\text{Am}_{mid}=1$. These simulations are initialised with random perturbations and are computed with  $L_x=L_y=20 H$, $L_z=8H$ and resolution $N_X=N_Y=256$, $N_Z=128$. The top panels of Fig.~\ref{fig_3d} show that in all cases, zonal flows are produced, with growth rates and size that can substantially vary from on case to another. To help the analysis, we show in the lower panels the evolution of the mean transport coefficients $\alpha_\nu$, $\alpha_L$, and mean mass loss efficiency $\zeta$. In the ideal limit (simulation without explicit diffusion), these quantities are almost identical to those obtained in 2D (see Fig.~\ref{fig_nonlinear} for comparison). However, the ring structure is much wider and takes longer time to form. The growth rate,  measured from the left panel of Fig.~\ref{fig_3d2} is indeed $\sigma_{3D} \simeq 0.0034 \simeq   \sigma_{2D}/2.5$. The main difference is that the turbulent magnetic diffusivity $\eta_t$ in 3D is considerably enhanced by the azimuthal (or non-axisymmetric) MRI dynamics. We indeed measure $\eta\simeq 0.12$ which corresponds to $Pm_t^{3D}  \simeq 2 \ll Pm_t^{2D}$, in agreement with other 3D numeric simulations \citep{lesur09b}. Therefore, according to the linear theory, the growth rate is reduced and drops rapidly with $k_x$ (unlike the 2D case for which it was quite flat.)  The maximum growth rate $\sigma \simeq 0.0039$ is obtained for $k_x H=0.2$ and stability is reached  for $k_xH \gtrsim  0.42$. This explains why we obtain a single ring in the 3D case, while 3 or 4 rings were obtained in 2D. We note that, unlike the 2D case,  $\alpha$ does not drop significantly after saturation of the instability and vigorous turbulent motions persist along with the zonal structure. \\

In non-ideal simulations,  Figure \ref{fig_3d} (bottom panels) indicates that the wind mass loss efficiency $\zeta \simeq 0.015$ is  comparable to that in the ideal case. However, the turbulent stress is weak or even reduced to zero in the ambipolar case.   The radial transport of angular momentum is provided essentially by the laminar component of the stress $-\langle B_x \rangle  \langle B_y \rangle$.  Azimuthal structures are very faint and have little impact on the unstable dynamics.  We checked indeed that turbulent diffusion of $B_z$ structures is negligible compared to the explicit one.   As a consequence, growth rates are much larger than in the ideal case. We measure respectively $\sigma=0.018$ and $\sigma=0.021$ in the ohmic and ambipolar simulations.  Using an extension of the linear theory (see  Appendix \ref{appendixB}) in the limit $\alpha_L \gg  \alpha_\nu$, we found that the theoretical growth rates match with the numerical values. \\

 Note that $\sigma$ is a factor 2 larger than in the 2D axisymmetric case. We attribute this difference to the fact that $\zeta$ is on average twice as large as in 2D.  Finally, with ambipolar diffusion, the number of rings is identical  between the 2D and 3D case. With pure ohmic diffusion, the 3D simulation exhibits a single ring (instead of 3 in 2D). It is possible that the stronger laminar stress  in 3D (a factor $\gtrsim 2$) favours the emergence of larger scale structures, thought initial conditions may also play a secondary  role in the final shape.

\subsection{Dependence on magnetization $\mu$}
Finally, we run a set of 3D ideal simulations (without explicit diffusion)  by varying $\mu$. We scan a large range of magnetizations from $\mueq=10^{-4}$ to $10^{-1}$.  For the largest magnetisations ($\mu=10^{-2}$ and $0.08$), we used a box of size $L_x=30$ and $40 H$ respectively. Two or three rings are obtained for $\mu=10^{-4}$ while a single ring is obtained for intermediate magnetizations.  In the extreme case ($\mu\sim0.08$), 3-4 rings form at the very early phase, but rapidly merge into a single ring. The growth rates of each prominent modes are shown in the right panel of Fig.~\ref{fig_3d2} (red diamonds). We superimpose the theoretical growth rate $\sigma(\mu)$ (purple line) obtained by using the linear theory of Section \ref{sec_theory}. Here we assume the same scaling relations for $\alpha$ and $\zeta$ as in 2D (Eq.~\ref{eq_scaling1}) but with $\eta_t=0.5 \alpha_\nu$.  The theory reproduces quite well the numerical values, except maybe in the large magnetization regime ($\mu\sim0.08$). In the same figure, we also plot the numerical growth rates obtained in the resistive run (yellow marker) and those measured in ambipolar runs (green markers). The green dashed line accounts for the growth rate computed with an extension of the linear theory, in the regime $\alpha_L\gg  \alpha_\nu$ (see Appendix \ref{appendixB}) assuming  $\zeta = 7 \, \mu^{0.85}$  and $\alpha_\nu=0; \alpha_L=4  \mu^{0.45}$. 
\section{Discussion and conclusion}
\label{sec_discussions}
In summary, we showed that the ring/gap structures obtained in  simulations of magnetized discs are formed via a linear instability. The process is generic and  works in all geometrical configurations (2D or 3D), with or without turbulence, and in various diffusive regimes  (ideal, ohmic or ambipolar). This instability is driven by a magnetic wind, associated with a large-scale poloidal field,  and relies on the assumption that the mass loss rate increases locally with the magnetization. The process works as follows:  
\begin{itemize}
\item A small radial perturbation of density generates a radial flow directed toward the gaps, via the turbulent stress if $\alpha\neq 0$.
\item The magnetic field, initially uniform, is radially transported towards the gaps
\item The excess of poloidal flux  induces a stronger wind and an excess of ejected mass in the gaps
\item The initial density perturbation is thus reinforced. 
\end{itemize}
We showed that in theory,  the instability exists in the "no-stress" regime ($\alpha=0$). {In that case, the radial flow results simply from  angular momentum conservation. Indeed, assuming a geostrophic balance and an initial density perturbation,  an azimuthal or "zonal"  flow is  produced.  Therefore, the inertial term in the azimuthal direction produces a radial flow opposed to the initial density ring.} However in practice the radial transport of matter and magnetic field is mainly (or even totally) induced by the viscous stress. Growth rates are typically of the order of the mass loss rate efficiency $\zeta=(\rho u_z)_{\mid_{z=H}}/(\Sigma\Omega^{-1})$. 
In Sections \ref{sec_simulations} and \ref{sec_simulations3D}, we brought several evidence that such linear instability is at work in our MHD simulations. We showed in particular that axisymmetric modes undergo exponential growth in the early phase of the simulations, at rates compatible with the linear theory.  A remarkable result is that growth rates increase rapidly with $\mu$, and can be of order $\simeq 0.1 \,\Omega$ for $\mu=0.1$. This is a direct consequence of the wind reinforcement at large $\mu$.  Although the instability can theoretically exist in the limit of small wavenumbers, the rings separation is set by non-ideal processes in the disc and is typically $\simeq 10H$ for realistic diffusion coefficients. The final ring density contrast can be also estimated from a simple saturation predictor.  Our model suggests that in typical T-Tauri discs with $\Sigma \propto R^{-1}$,  the dust accumulates into the gas rings for magnetization $\mu \gtrsim 10^{-4}$. We finally provided detailed diagnostics of the axisymmetric  flow, which are all consistent with the instability mechanism described above. \\

Unlike the mechanism suggested  by \citet{lubow94,cao02}, the magnetic torque exerted by the wind is unnecessary here. We note that our linear analysis has been carried out around antisymmetric equilibria, for which the mean vertical stress and accretion flow are zero. Given its robustness, we think that the inclusion of a mean wind torque in our model will not dramatically affect the process, though an extension of the linear theory needs to be developed in this configuration. One possible effect is that the ring structures will be advected by the mean accretion flow and therefore slowly drift toward the central object.  Global simulations with ambipolar diffusion have shown that ring structures persist despite the recurrent change in the large scale wind geometry and vertical symmetries during the disc evolution \citep{bethune17,suriano18b}. In any case, further investigations will be  necessary to characterize the instability mechanism in the global configuration.{Another caveat is related to the effect of boundary conditions and mass replenishment. Indeed, the mass loss rate is known to depend on the location of the vertical boundary \citep{fromang13}, as the outflow never crosses the fast magnetosonic point. The growth rate is then impacted too, according to our calculations. We however expect then the dependence of growth rates on $\mu$ is still correct, but that the absolute values of $\sigma$ might depend on the external environment of the disc, but also on the topology of magnetic fields and wind launching conditions, which in reality can differ from the simulations and from object to objects. Regarding the mass replenishment procedure,   it takes place on a long timescale, similar to the instability timescale. We however think that it does not affect the whole process since it is independent on the radial coordinate. We also conducted simulations without replenishment and found that rings still form within the same timescale, despite the progressive loss of mass experienced by the disc. This indicates that mass replenishment does not play any role in the ring formation mechanism. }
\\

We think that the process described in this paper could have important implications for discs around young stars.  Observationally, there are  indications that some discs may emit a wind  through the action of a large scale poloidal field. Indeed recent studies have reported low-velocity molecular outflow consistent with MHD winds  in Class 0 and I discs \citep{Launhardt09,Bjerkeli16,tabone17,hirota17,louvet18}, although alternative processes are not excluded (photo-evaporation winds or jet "cocoon"). In theory, the magnetic flux from the parental core during the disc formation could be sufficient to provide the degree of disk magnetization required for the launching of a wind.  However an important unknown is how the poloidal magnetic flux is radially transported during the disk lifetime. If such transport is efficient, as suggested by recent works  \citep{guilet13,zhu18},  the  wind instability  process described in this paper may be possible during the early stages of the disc evolution (Class 0 to Class II).  As shown in Section \ref{sec_simulations}, the instability form long-lived rings of size $\gg H$, which might remain stable during a large fraction of the disc lifetime. Such wind-driven instability could potentially explain the gaps  imaged by ALMA or SPHERE in young discs like HL Tau or eventually leads to cavities like those observed in transitional discs \citep[an alternative is the transonic wind-driven accretion recently proposed by][]{wang17}. 

Recent  MHD simulations with similar setup  \citep{riols18}  have shown that the pressure maxima associated with the rings can efficiently trap dust grains of intermediate size (mm to dm). This could be directly relevant for planet formation theory, as it may help to overcome  the "radial-drift" barrier  (growth of grains to pebbles)  and the "fragmentation" barrier ( growth from planetesimals to planets)   during the early phases when dust abounds  \citep{birnstiel16}. Indeed, the confinement of solids in pressure maxima 
 can stop the radial migration of grains, accelerate their growth, and prevent their fragmentation \citep{gonzalez17}.  Further work is however needed to understand how fast grain can grow and how the dust back-react on the gas rings. Another interesting avenue of research is to investigate  the interaction between these rings and the disc gravitational instability, which is  expected in class 0  and class I  discs 
\citep{tobin13, mann15}. It is indeed unclear whether self-gravity could enhance or destroy the zonal MHD flows. 

Finally, the linear instability identified in this paper could be applied to other accreting systems, such as Active Galactic Nuclei, dwarf novae or X-ray binaries. Indeed some of these objects show signatures of strong jets, potentially driven by large scale poloidal fields via a magnetocentrifugal effect. The spontaneous formation of rings via a linear instability could have important consequences in their dynamics and their variability. 

\begin{acknowledgements}
This work acknowledges funding from The French ANR under contracts ANR-17-ERC2-0007 (MHDiscs). This work was granted access to the HPC resources of IDRIS under the allocation  A0040402231 made by GENCI (Grand Equipment National de Calcul Intensif). Part of this work was performed using the Froggy platform of the CIMENT infrastructure (https://ciment.ujf-grenoble.fr), which is supported by the Rhône-Alpes region (GRANT CPER07-13 CIRA), the OSUG@2020 labex (reference ANR10 LABX56) and the Equip@Meso project (reference ANR-10-EQPX-29-01) of the programme Investissements d'Avenir supervised by the Agence Nationale pour la Recherche.
\end{acknowledgements}



\bibliographystyle{aa}
\bibliography{refs} %if your bibtex file is called example.bib

\appendix 

\section{Linearisation of total stress in the azimuthal momentum equation}
\label{appendixA}
In this appendix, we linearise the total stress in the $y$-momentum equation (radial plus vertical) around a given disc equilibrium. For that we use assumptions (2) and (3) of Section \ref{linearisation}.  
These assumptions are checked numerically for unstable modes around a laminar disc equilibrium (see appendix \ref{appendixC}). Using the formalism of Section \ref{sec_theory} and Eq.~(\ref{eq_Txy}), Eq.~(\ref{eq_Wyz}), the linearised quantities are
\begin{equation}
\Sigma \overline{T}_{yx}= \Sigma_0  \alpha_{\nu_0}\left(\hat{\Sigma}+ q \hat{\mu}\right) c_s^2 -  B_{x_0} \underline{B_{y_0}} \hat{b}_y -  B_{y_0}\underline{ B_{x_0} }\hat{b}_x
\end{equation}
\begin{equation}
W_{yz} = -\underline{B_{y_0}} B_{z_0} ( [\hat{b}_z]^+_- +[\delta \hat{b}_y]^+_-)
\end{equation}
Using the solenoidal condition , integrated azimuthally and vertically, we obtain that $ B_{z_0}  [\hat{b}_z]^+_- = - i k_x \underline{ B_{x_0} }\,\hat{b}_x$. Therefore, the term associated with the vertical stress perturbation compensates exactly the term $- B_{y_0} \underline{B_{x_0}} \hat{b}_x$ in the laminar radial stress. Adding $W_{yz}$ and the $x$ derivative of the averaged radial stress gives: 
\begin{multline}
\dfrac{\partial}{\partial x}\left ({\Sigma \overline{T}_{yx}}\right)+ {W}_{yz}  =  i k_x \Sigma_0  c_s^2 \left[\alpha_{\nu_0}\left(\hat{\Sigma}+ q \hat{\mu}\right)+ \alpha_{L_0} \hat{b}_y\right] \\- ik_x B_{z_0} \underline{ B_{y_0} } [ \hat{b}_y]^+_-
\end{multline}
with $\alpha_{L_0} = -\underline{B_{x_0}} B_{y_0}/(\Sigma_0 c_s^2)$. We need then to relate the perturbation $\hat{b}_y$ to the variables of the reduced model $\hat{\Sigma},  \hat{\mu}$ or  $\hat{u}$. For that, we assume that vertically,  the disc is in a magneto-hydrostatic equilibrium and the magnetic field is essentially toroidal:  
\begin{equation}
\label{eq_hydrostat}
\dfrac{\partial}{\partial z} \left(\rho c_s^2 + \dfrac{B_y^2}{2}\right) = - \rho  z \Omega^2
\end{equation}
This is true below the Alfven point, where inertial terms of the wind are small compared to magnetic pressure.  If we integrate Eq.~(\ref{eq_hydrostat}) between $-z_d$ and $z_d$, we obtain immediately that any perturbation of $B_y$ is symmetric with respect to the midplane  $[\hat{b}_y]^+_-=0$ (if the background equilibrium $B_y$ is also symmetric). We now integrate Eq.~(\ref{eq_hydrostat}) between $z_i$ and $z$, with $z_i$ chosen to have $B_y^2 [z_i]= \langle \underline{\delta B_y^2} \rangle$. This altitude is always defined since $B_y$ is a decreasing function of $z$ and  is located in general below the Alfven point since $B_y$ is close to 0 at this altitude.  By averaging in $z$ and assuming $\rho[z_i]\approx 0$, we obtain then
\begin{equation}
\langle \rho c_s^2 \rangle  + \dfrac{\langle B_y \rangle\langle B_y \rangle^\ast}{2} = \left \langle \int_{z_i}^z - \rho (z') z' \Omega^2 dz \right \rangle 
\end{equation}
If we assume that the density profile is not too far from a Gaussian, of width $H_m  = h H$ with $h>1$, then we have: 
\begin{equation}
\int_{z_i}^{z}  \rho(z')  z' \Omega^2 dz' \simeq -\rho\,h^2 c_s^2
\end{equation}
and we find
\begin{equation}
 {\langle B_y \rangle \langle \underline{B_y} \rangle}   \simeq  2 (h^2-1)  \Sigma  c_s^2
\end{equation}
We linearised this relation and obtain the following relation  for the dimensionless toroidal perturbation :
\begin{equation}
\hat{b}_y  \simeq  \lambda \hat{\Sigma}    \quad \text{with}  \quad  \lambda = \dfrac{1}{2}\dfrac{h^2-1}{\mu_{y_0}}\simeq 0.5
\end{equation}
$\mu_{0_y}= B_{y_0}^2/(2\Sigma c_s^2)$ is the "toroidal magnetization". The total linearised stress can be then written: 
\begin{equation}
\dfrac{\partial}{\partial x}\left ({\Sigma \overline{T}_{yx}}\right)+ {W}_{yz} = i k_x  \left [ \left (\alpha_{\nu_0} + \lambda \alpha_{L_0}\right) \hat{\Sigma} +  \alpha_{\nu_0}  q \hat{\mu}\right]  \Sigma_0  c_s^2
\end{equation}
\section{Extension of the linear theory to laminar discs or discs with strong toroidal field}
\label{appendixB}
In this appendix, we relax assumptions (1) and (3) in Section \ref{linearisation} and extend the linear analysis to a more general configuration. The geostrophic balance is replaced by a magneto-geostrophic balance, in which we include the effect of toroidal magnetic pressure. The radial stress can be purely laminar or a mix of both turbulent and laminar components.  The following calculation applies in particular to discs with ambipolar diffusion. The linearised system of equations is:
\begin{equation}
\left (
\begin{array}{ccc} 
\sigma+\zeta_0 & ik_x &  p\zeta_0 \\
\dfrac{\sigma}{2} (1+\mu_{y_0}\lambda) +\alpha_{\nu_0}+\lambda\alpha_{L_0} & -\dfrac{i}{2k_x} & \alpha_{\nu_0}q\\
\sigma+\eta^\star k_x^2 & 2ik_x& \sigma+\eta^\star k_x^2  \\
\end{array}
\right)  \left (\begin{array}{c} \hat{\Sigma} \\\hat{u}\\ \hat{b}_z \end{array} \right) = 0
\end{equation}
with $\lambda \approx 0.5$, a parameter introduced in Appendix \ref{appendixA} and $\mu_{0_y}= B_{y_0}^2/(2\Sigma c_s^2)$ the toroidal magnetization of the equilibrium. The growth rates follow the dispersion relation
\begin{equation}
A \sigma^2 + B\sigma + C = 0
\end{equation}
with 
\begin{multline}
A = 1+ k_x^2(1+\mu_{y_0}\lambda)  \\
B=   2k_x^2 \left[\alpha_{\nu_0}(1+q)+\lambda\alpha_{L_0}\right] - \zeta_0 \left [p-1+2pk_x^2(1+\mu_{y_0}\lambda)\right] \\
+\eta^\star \left[k_x^2+k_x^4(1+\mu_{y_0}\lambda)\right]\\
C=-   4  \zeta_0 k_x^2\left[\alpha_{\nu_0}(p-q)+p\lambda\alpha_{L_0}\right]\\  + \eta^\star k_x^2 \left[\zeta_0(1-p)+2\alpha_{\nu_0}(1-q)k_x^2+2\lambda \alpha_{L_0}k_x^2\right]\\
\end{multline}
Note that the case with a pure laminar stress is equivalent to that of a pure turbulent stress, but with $q=0$. Thus, the instability criterion for laminar disc is less restrictive. In practise, given the fact that $q=0$ and that turbulent diffusion is absent,  growth rates can be enhanced (by a factor $\approx  2$) compared to the turbulent case. A strong toroidal field can also help the instability, by affecting the  geostrophic equilibrium  and enhancing the zonal flows. 
\section{Stability of laminar disc with initial 1D uniform wind}
\label{appendixC}
\begin{figure}
\centering
\includegraphics[width=\columnwidth]{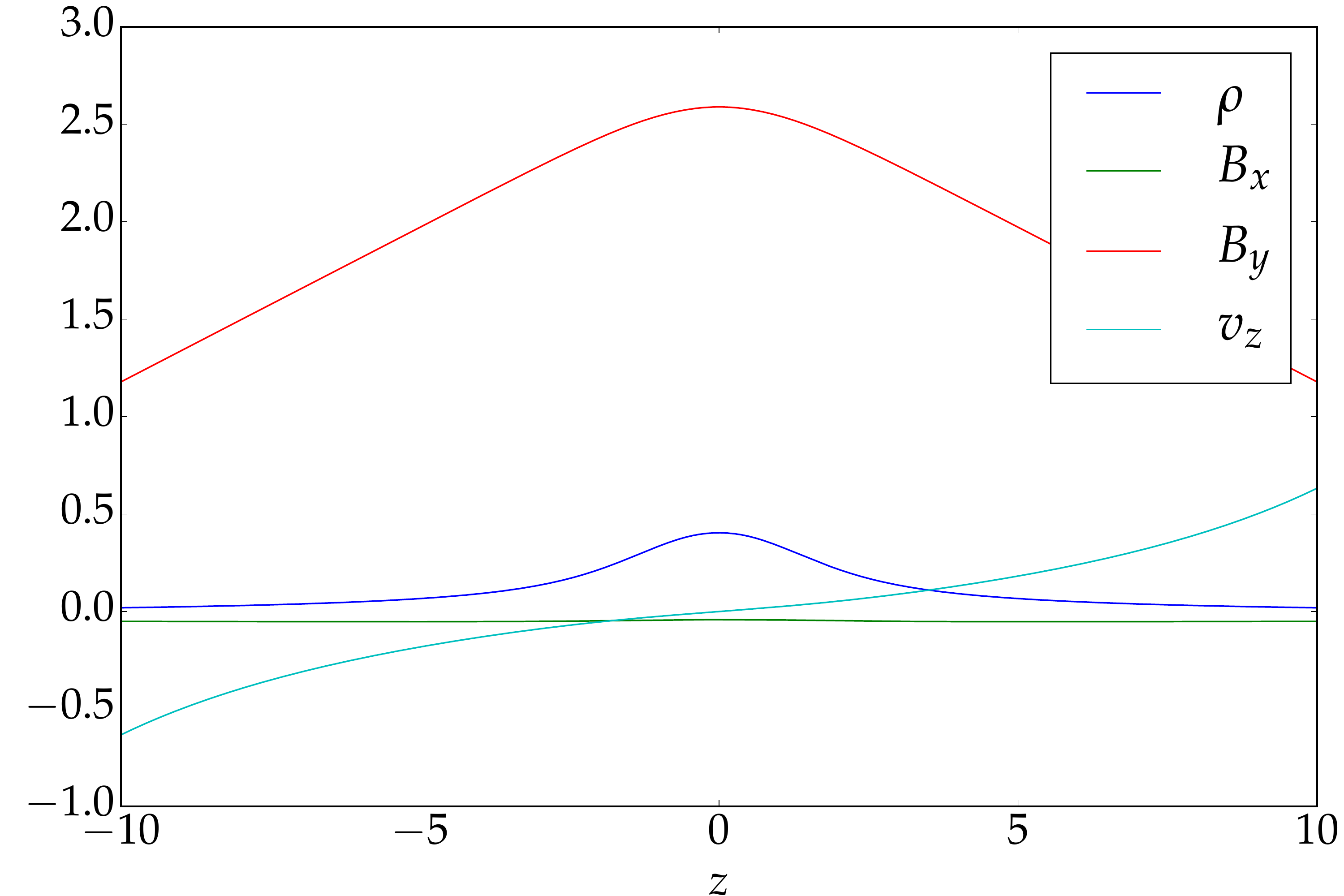}
\includegraphics[width=\columnwidth]{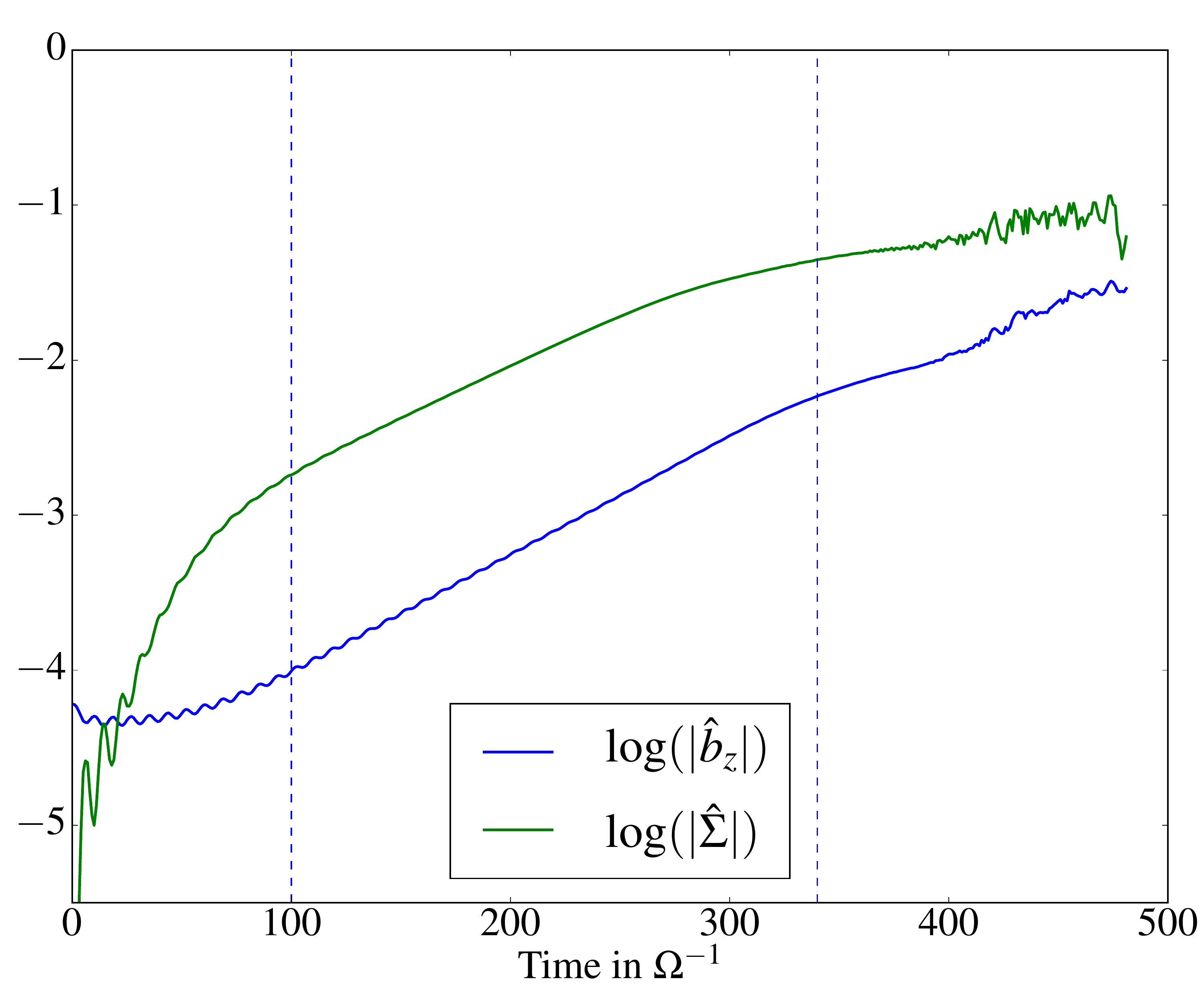}
 \caption{Top: Some quantities related to the 1D wind equilibrium. Bottom: Amplitude (in logarithm) of the unstable Fourier mode $k_x=\pi/10$ as a function of time. The background equilibrium is a laminar 1D wind computed for $ \mueq=10^{-3}$. Blue and green lines are respectively the surface density and vertical magnetic field perturbation. The blue dashed lines delimits the linear growth phase.}
\label{fig_linear_growth}
\end{figure}
\begin{figure}
\centering
\includegraphics[width=\columnwidth]{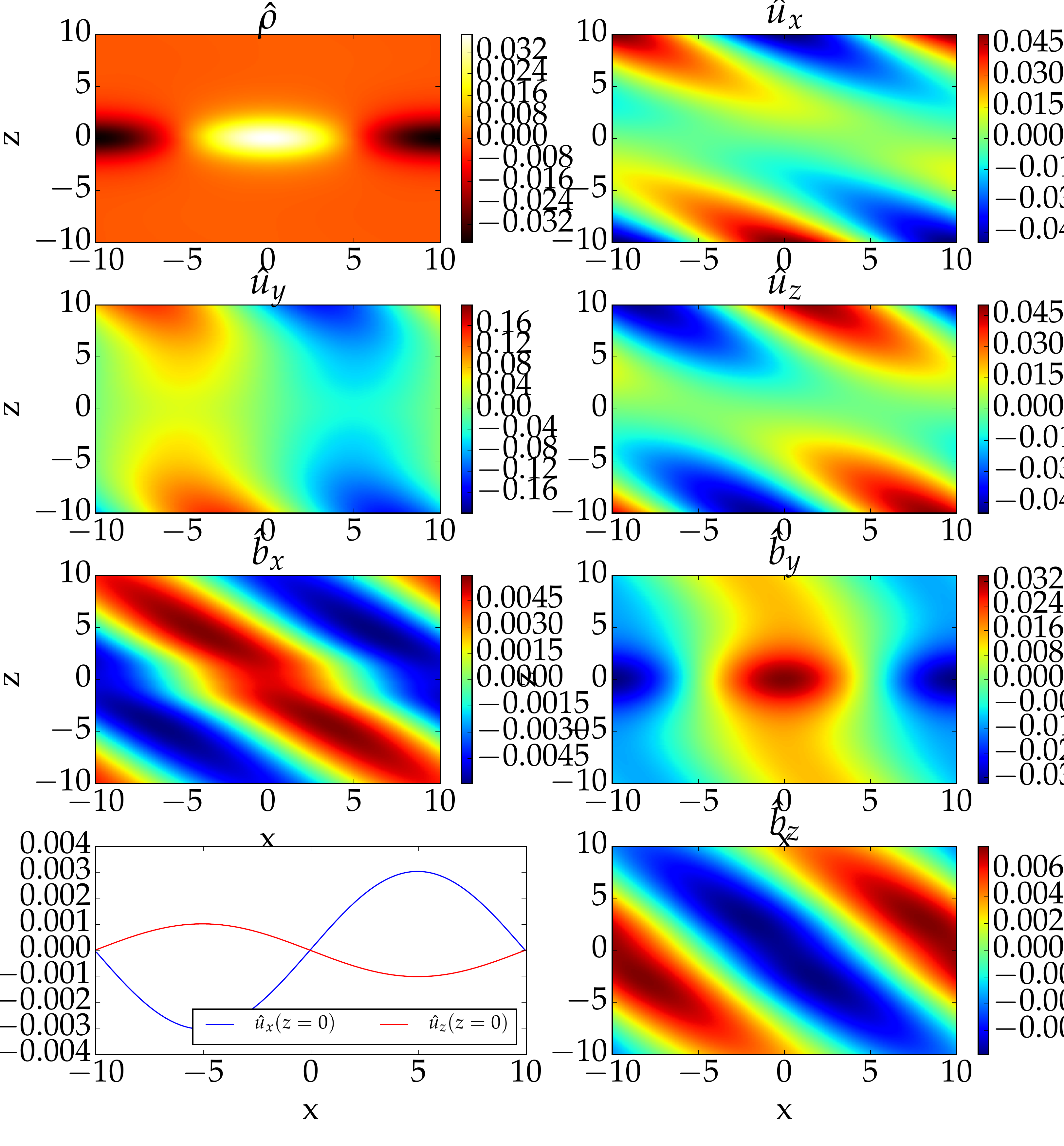}
 \caption{Shape, in the poloidal plane, of the unstable axisymmetric mode $k_x=\pi/10$ around the 1D wind equilibrium ($ \mueq=10^{-3}$). }
\label{fig_laminar_umode}
\end{figure}
\begin{figure}
\centering
\includegraphics[width=\columnwidth]{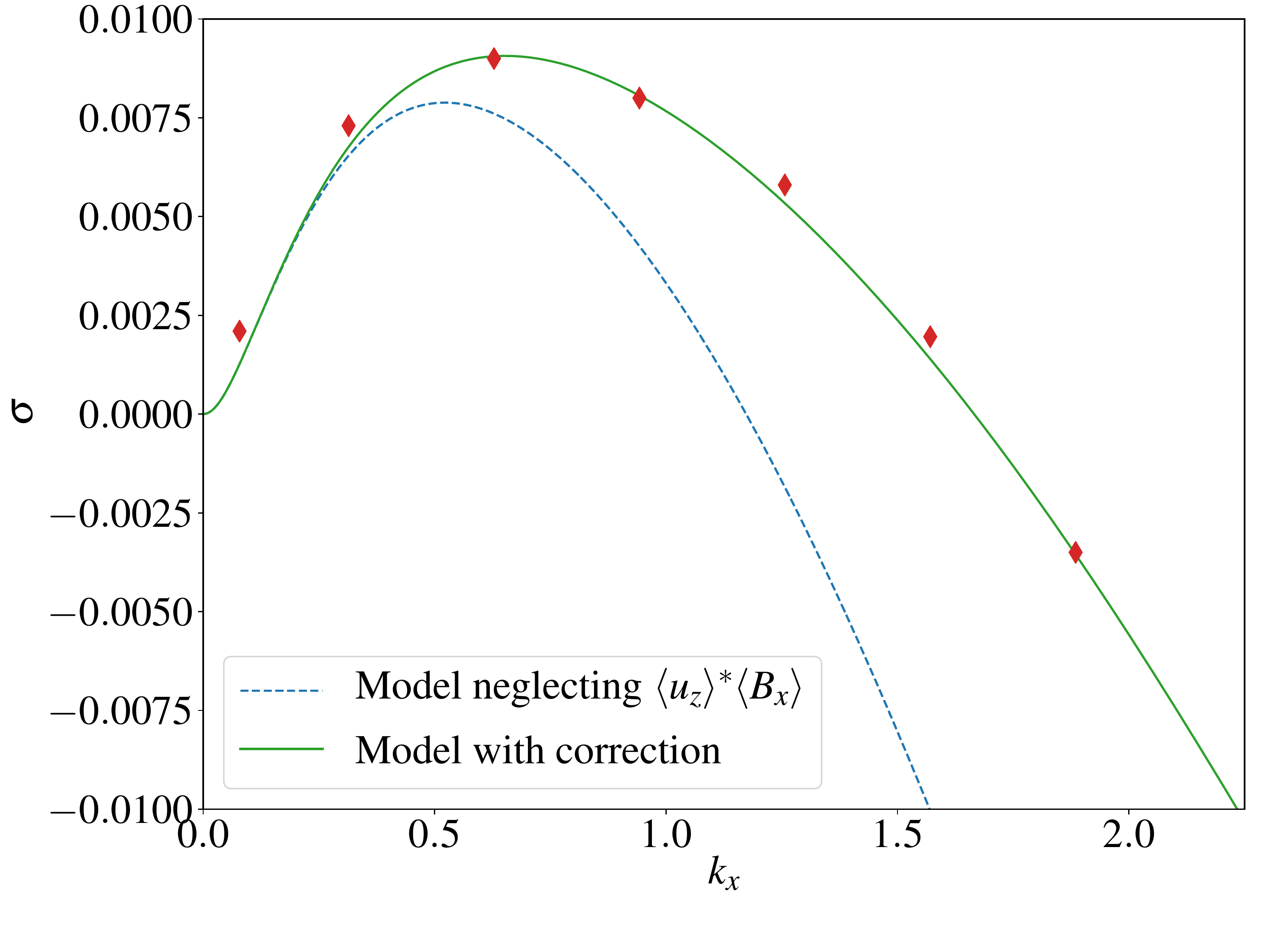}
 \caption{Growth rate of axismmyetric modes measured in the laminar-wind simulations for different $k_x$ (red diamond makers). The blue-dashed curve is the theoretical growth rate predicted from the model extension in Appendix \ref{appendixB} (assuming a pure laminar stress). The green plain curve is the same model but taking into account the $u_z B_x$ term in the induction equation (this term is simply estimated by calculating the ratio  $u_z B_x/u_x B_z$ associated with the unstable mode in simulations.}
\label{fig_growthrate_laminar}
\end{figure}

To rigorously demonstrate the existence of an instability, similar to that predicted in Section \ref{sec_theory}, we use a very simple setup which consists of an initial uniform and laminar wind solution (independent on $x$).  In that case, the radial stress is provided by the mean $-B_x B_y$ relative to the laminar solution. 

The first step is to compute a wind equilibrium in the local frame. For that, we employ the same technique as \citet{lesur13}: we run a 1D shearing box simulation with an initial hydrostatic disc  threaded by a vertical field of strength $B_{z_0}=0.044$ ($\mueq=10^{-3}$). We add some random noise at $t=0$ and let the system evolves until it reaches a new equilibrium. In-between, an MRI mode is triggered and lead to the  launch of a magneto-centrifugal wind. Figure \ref{fig_linear_growth} (top) shows the equilibrium obtained in a box that spans -15 to $15 H$ in the vertical direction. {This class of solution is characterized by a very strong toroidal field and develops only if the system is initialised with a very clean 1D perturbation, so it is likely that it is never encountered in nature.} The wind is however more realistic with mass loss efficiency $\zeta_0 \simeq 0.01$. 

We then run a 2D axisymmetric simulation, starting from the wind solution and add a small perturbation of the form $\exp(ik_x x)$ with $k_x=2\pi/L_x$. Here we use a box size $L_x=20 H$ and resolution $(N_X,N_Z)=(256,512)$.  Figure \ref{fig_linear_growth} (bottom) shows the time-evolution of the axisymetric perturbation $\hat{b}_z$ and $\hat{\Sigma}$, projected onto the Fourier component $k_x=2\pi/L_x$. 
Clearly, the mode is amplified exponentially during the first 200 $\Omega^{-1}$ with growth rate $\sigma\simeq 0.0074$.  We show in Fig.~\ref{fig_laminar_umode} the shape of the unstable mode in the poloidal ($x$,$z$) plane. The density field develops a ring structure while the magnetic perturbation,  anti-correlated with $\rho$,  forms an inclined shell similar to that obtained in 3D turbulent simulations (see Fig.~\ref{fig_ring_beta3Am1} for comparison). The radial velocity  of the mode is anti-correlated with $\hat{v}_y$ and directed toward the gap. The outflow and mass loss rate perturbations $\rho_{0} \hat{u}_z$ are positive inside the gap and negative outside. All of these properties are indications that the mode is triggered by the same instability as described in Section \ref{sec_theory}. One other interesting result is the apparent correlation between the density and the $B_y$ structures, near the midplane. Such behaviour results from the magneto-hydrostatic equilibrium in the vertical direction (very well checked) and actually confirms our calculation in Appendix \ref{appendixA}. In particular we found $\lambda = \hat{b}_y/\hat{\Sigma}\simeq 0.2$ when quantities are integrated vertically up to the Alfven point  (instead of $\lambda\simeq 0.5$ in Appendix \ref{appendixA}). This parameter does not seem to depend on the integration boundary neither on $k_x$. 

We finally achieve the same simulation for different $k_x$. The growth rates obtained for seven different radial wavenumbers are shown in Fig.~\ref{fig_growthrate_laminar} (red diamond markers). The blue dashed curve is the model prediction, assuming $\alpha_L=0.4$,  $\zeta_0=0.01$, $\lambda\simeq 0.2$, $\lambda \mu_{0y}$= 3.2 and $p=0.5$. Note that we use the extended model of Appendix \ref{appendixB} to compute the growth rates. The value of $p$ is different from turbulent simulations and has been obtained by varying the  magnetization and measuring $\zeta$ of different 1D wind solutions. Although the maximum growth rate is in agreement with the numerical data, the behaviour at large $k_x$ is not well reproduced by our model.  The reason is the omission of the EMF term $u_z b_x$ in the induction equation, which clearly enhances the instability at large $k_x$. We measured the ratio $u_z b_x/u_x b_z$ for each $k_x$ and include in the model a correction factor $c_o(k_x)$ in the linearised induction equation. With such correction, we obtain a very good match between the numerical and theoretical growth rates at large $k_x$ (plain green curve in Fig~.\ref{fig_growthrate_laminar}). 
\section{Check of model assumptions in simulations}
\label{appendixD}
\begin{figure*}
\centering
\includegraphics[width=\textwidth]{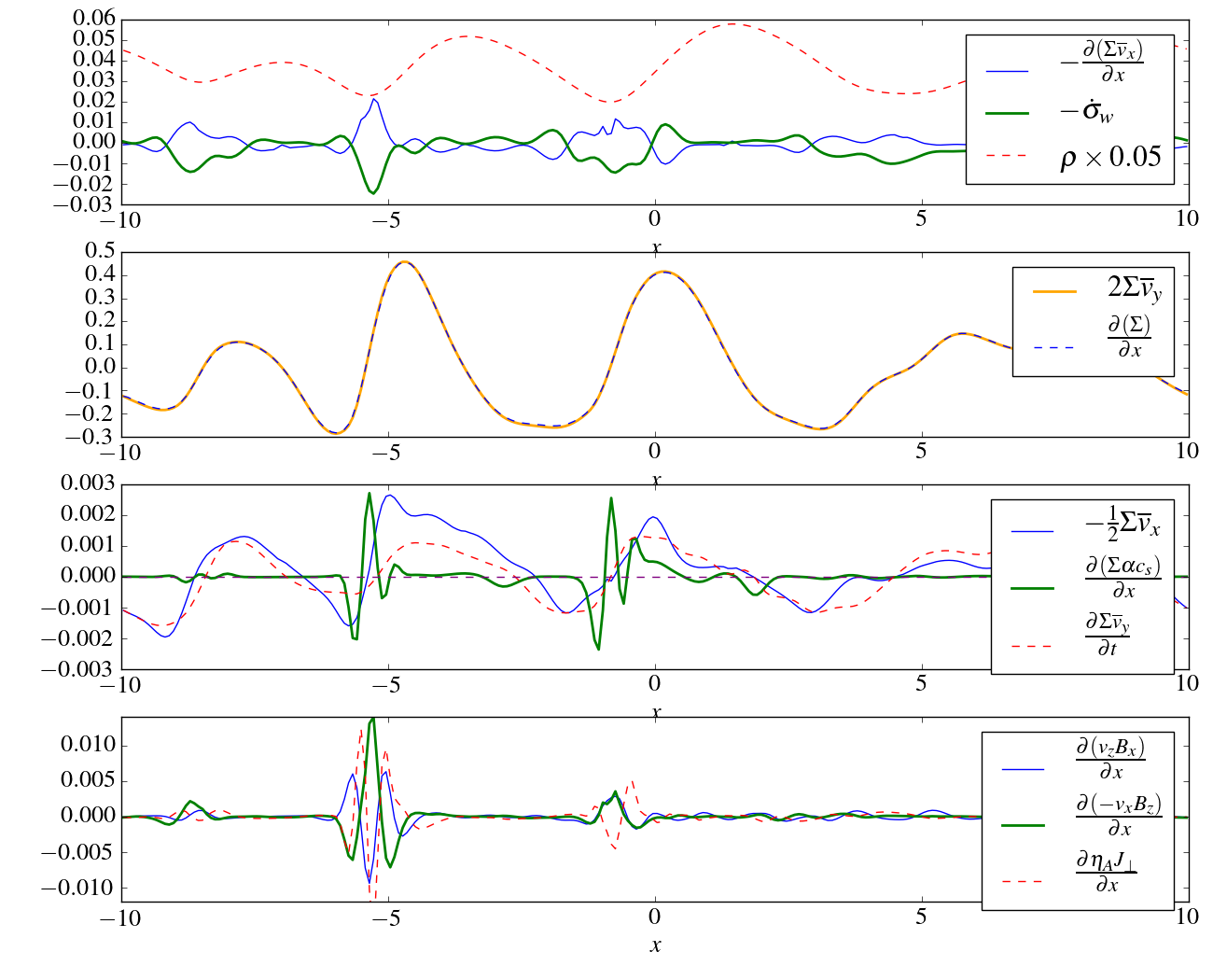}
 \caption{Radial profiles of the flux and source terms in equations \ref{mass_eq}, \ref{mx_eq},  \ref{my_eq}  and \ref{bz_eq}, integrated vertically within $z\pm H$ and averaged in time during the linear phase for $\mueq=10^{-3}$ and $\text{Am}=1$.  The linear phase is delimited by the vertical dashed  lines in Fig.~\ref{fig_2Dambipolar} ($60 <t<  200 \, \Omega^{-1}$.)}
\label{fig_forcebalance}
\end{figure*}
\begin{figure}
\centering
\includegraphics[width=\columnwidth]{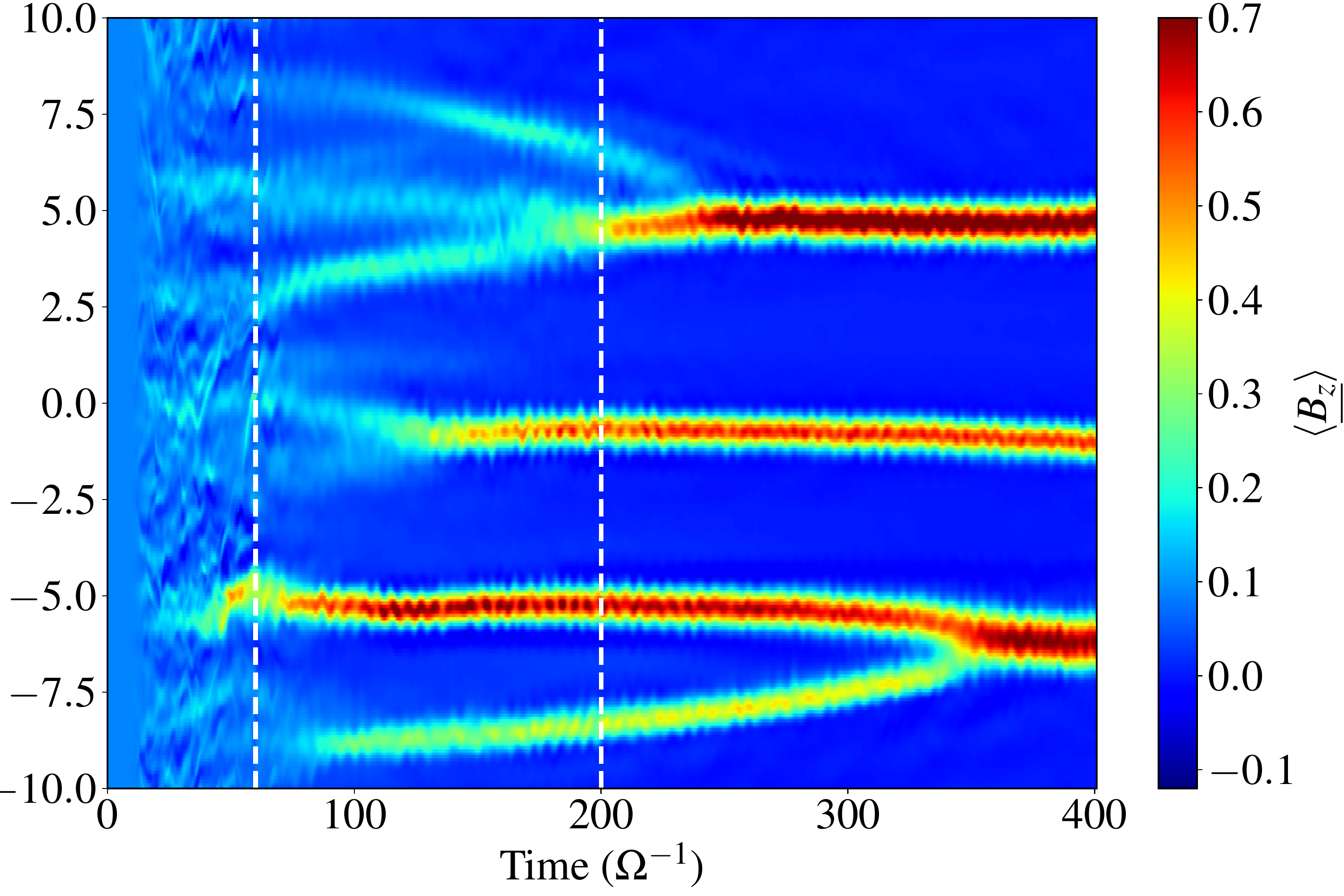}
 \caption{Spacetime diagrams showing the vertically averaged $B_z$  in the 2D ambipolar simulation ($\text{Am}_{mid}=1$,  $\mueq=10^{-3}$ and  $z_d=2 H$). The two vertical dashed lines delimit the "linear" phase, during which zonal modes, in Fourier space,  grow exponentially. }
\label{fig_bz2}
\end{figure}
To check the assumptions made in Section \ref{linearisation} and validate the linear theory of Section \ref{sec_theory}, we examine in details the  mass, momentum and magnetic budget in the 2D ambipolar run with $\mu=10^{-3}$. Figure \ref{fig_forcebalance} shows the
 different flux and source terms in equations (\ref{mass_eq}), (\ref{mx_eq}),  (\ref{my_eq})  and (\ref{bz_eq}), averaged during the growth phase (delimited by the dashed lines in Fig.~\ref{fig_bz2}). The terms are calculated in real space (and thus comprise all the axisymmetric modes) and are integrated within $\pm  1 H$. The first panel shows that the gaps are depleted  by the vertical mass flux $\dot{\sigma}_w$ while they are re-filled by the radial mass flux. It can be noted that the first term is slightly larger than the second.  This ensures that the instability is driven by the outflow. The second panel clearly demonstrates that perturbations are in a geostrophic balance (assumption 1 in Section \ref{linearisation}), the Coriolis force being equilibrated by the radial thermal pressure gradient. In the azimuthal direction (third panel),  both the radial flux of turbulent stress and the inertial term $\sim \sigma \Sigma_0 \overline{v}_y$ contribute to the production of $v_x$, directed toward the gaps.  As suggested by the linear equations in Section \ref{linearisation}, the inertial term due to the zonal flow can be an additional source of radial transport and even substitute for the $\alpha$ viscosity. 

Finally, in the last panel, we show that the vertical field in the gaps is mainly produced via the  EMF term associated with radial motion -$v_x B_z$, as expected from the theory.  The role of the vertical EMF component $v_z B_x$ is more ambiguous: it seems to reinforce $B_z$  in the nascent gaps at $x\simeq-8H$ and $x=0$ but has a negative effect in the central gap at $x=-5$. Actually, Fig.~\ref{fig_bz2} shows that this structure has already evolved into a non-linear regime, unlike the others. In any case, this EMF term can be comparable to  -$v_x B_z$ and thus, our assumption (4) in Section \ref{linearisation} is not necessarily satisfied. The ambipolar term $ \eta_A J_{\perp_y}$ (with $J_{\perp_y}$ the $y$ projection of the  current perpendicular to $\mathbf{B}$),  is always negative within the gaps and positive outside. Therefore it tends to diffuse the magnetic field out of the gaps. This last result is clear evidence that the process forming rings (or gaps) is supported by the ideal terms and not by the ambipolar diffusion. In conclusions, we showed that the force balance assumed in the linear regime (Section \ref{linearisation}) is compatible with numerical data. 
\end{document}